\documentclass[11pt]{article}
\usepackage{hyperref}
\usepackage{graphicx}
\usepackage{graphics}
\usepackage{dsfont}
\usepackage{epsfig}
\usepackage{bbm}
\usepackage{xypic,mathtools}
\usepackage{amsmath,amssymb,amsthm,amscd}
\usepackage[sort&compress,square,numbers]{natbib} 
\usepackage{chemarrow}
\usepackage{float}
\usepackage{scalefnt}
\usepackage{wrapfig}
\usepackage{fancybox}
\usepackage{ifsym}
\usepackage{enumitem}
\usepackage{relsize}
\usepackage{multicol}
\usepackage{multirow}
\usepackage[small]{caption}
\usepackage{capt-of}

\usepackage{url}
\usepackage{hyperref}

\usepackage{lscape}

\usepackage{blkarray}
\usepackage{braket}
\restylefloat{table}

\usepackage{tikz}
\usetikzlibrary{calc,fit,matrix}

\usepackage{listings}

\usepackage{arydshln}
\usepackage[all,cmtip]{xy}

\def\clap#1{\hbox to 0pt{\hss#1\hss}}

\def\mclap{\mathpalette\mathclapinternal}

\def\mathclapinternal#1#2{%
\clap{$\mathsurround=0pt#1{#2}$}}

\def\Mgn[#1]#2{{\overline{\cal M}_{#1,#2}}}
\def\pqs[#1,#2]{{\footnotesize{$\left[\begin{array}{c} #1\\#2  \end{array}\right]$}}} 
\def\pqsu[#1,#2]{\left[\begin{array}{c} #1\\#2  \end{array}\right]} 
\def\pqssu[#1,#2]{{\footnotesize{\left[\begin{array}{c} #1\\#2  \end{array}\right]}}} 
\def\pqh[#1,#2]{{\footnotesize{$\left[\begin{array}{c} #1\\#2  \end{array}\right]$}}} 
\def\pqhu[#1,#2]{\left[\begin{array}{c} #1\\#2  \end{array}\right]}

\newcommand{\ba}{\begin{eqnarray*}}
\newcommand{\ea}{\end{eqnarray*}}
\newcommand{\ban}{\begin{eqnarray}}
\newcommand{\ean}{\end{eqnarray}}
\newcommand{\be}{\begin{equation}}
\newcommand{\ee}{\end{equation}}
\newcommand{\ben}{\begin{equation}}
\newcommand{\een}{\end{equation}}
\numberwithin{equation}{section}

\newcommand{\sequence}[1]{\ensuremath{\stackrel{\text{#1}}{\longrightarrow}}}

\numberwithin{equation}{section}

\newcommand{\bea}{\begin{eqnarray}\displaystyle}
\newcommand{\eea}{\end{eqnarray}}

{\setlength{\fboxsep}{15pt}
\setlength{\mylength}{\linewidth}%
\addtolength{\mylength}{-2\fboxsep}%
\addtolength{\mylength}{-2\fboxrule}%
\Sbox
\minipage{\mylength}%
\setlength{\abovedisplayskip}{0pt}%
\setlength{\belowdisplayskip}{0pt}%
\equation}%
{\endequation\endminipage\endSbox
\[\fbox{\TheSbox}\]}

\usetikzlibrary{decorations.pathmorphing}

\usepackage{float}

\def\2{{1\over2}}
\let\<=\langle \let\>=\rangle
\def\new#1\endnew{{\bf #1}}
\def\ifundefined#1{\expandafter\ifx\csname#1\endcsname\relax}
\ifundefined{draftmode}\else    \input draftmode        \fi
\let\Msize=\footnotesize             
\def\BM{\Msize\begin{matrix}}           \def\EM{\end{matrix}}
\def\MN M:#1 #2 N:#3 #4 {{(#1_{#2},#3_{#4})}}
\def\MNH M:#1 #2 N:#3 #4 H:#5,#6 [#7]{{(#1_{#2},#3_{#4})^{#5,#6}_{#7}}}

\def\r{\mathrm{r}}

\begin{document}

\begin{titlepage}
{}~ \hfill\vbox{ \hbox{} }\break

\rightline{
BONN-TH-2017-09}

\vskip 3 cm

\centerline{\Large \bf Modular Amplitudes and Flux-Superpotentials}\vskip 0.5 cm
\centerline{\Large \bf on elliptic Calabi-Yau fourfolds}

\renewcommand{\thefootnote}{\fnsymbol{footnote}}
\vskip 30pt \centerline{ 
Cesar Fierro Cota\footnote{fierro@th.physik.uni-bonn.de}, Albrecht Klemm\footnote{aklemm@th.physik.uni-bonn.de}, Thorsten Schimannek\footnote{schimann@th.physik.uni-bonn.de} } \vskip .5cm \vskip 20pt

\begin{center}

{$^{* \dagger\ddagger}$Bethe Center for Theoretical Physics and $\dagger$Hausdorff Center for Mathematics,}\\ 
{Universit\"at Bonn, \  D-53115 Bonn}\\ [3 mm]
\end{center}

\setcounter{footnote}{0}
\renewcommand{\thefootnote}{\arabic{footnote}}
\vskip 60pt
\begin{abstract}
We discuss the period geometry and the topological string amplitudes on elliptically fibered 
Calabi-Yau fourfolds in toric ambient spaces. In particular, we describe a general procedure to fix 
integral periods.
Using some elementary facts from  
homological mirror symmetry we then obtain Bridgelands involution and its monodromy 
action on the integral basis for non-singular elliptically fibered fourfolds. The full  monodromy group contains  
a subgroup that acts as PSL(2,Z) on the K\"ahler modulus of the fiber and we analyze the consequences of this modularity for the genus 
zero and genus one amplitudes as well as the associated  geometric invariants.
We find holomorphic anomaly equations for the amplitudes, reflecting  precisely  the failure of exact PSL(2,Z) invariance that relates them to quasi-modular forms. 
Finally we use the integral basis of periods to study the horizontal flux superpotential and the leading order K\"ahler potential 
for the moduli fields in F-theory compactifications globally on the complex structure moduli space. 
For a particular example we verify attractor behaviour at the generic conifold given an aligned 
choice of flux which we expect to be universal.
Furthermore we analyze the superpotential at the orbifold points but find no stable vacua.
\end{abstract}

\end{titlepage}
\vfill \eject

\newpage

\baselineskip=16pt    

\tableofcontents

\section{Introduction}

At present F-theory compactifications on elliptic Calabi-Yau 
fourfolds provide the richest class of explicit $N=1$ 
effective theories starting from string theory. The reason is 
that the construction of Calabi-Yau fourfolds as algebraic 
varieties in a projective ambient space is very simple and 
toric, or more generally non-abelian gauged linear $\sigma$-model 
descriptions provide immediately trillions of geometries~\cite{Kreuzer:2009}.

In fact, geometric classifications of certain compactifications 
with restricted physical features seem possible even though this 
has been achieved mostly for elliptic Calabi-Yau threefolds, 
where it has been argued that there exists only a finite 
number of topological types in this class~\cite{MR1272978}.

Most of the generic compact toric examples allow for elliptic fibrations and in addition 
for each of them there is a huge degeneracy of possible flux choices, 
which together with non-perturbative effects have been argued to 
solve the moduli stabilization problem by driving the theory to 
a particular vacuum. Ignoring the details of how this happens for the 
concrete geometry under consideration it has been shown that 
by degenerating the fourfold in a controlled way viable 
phenomenological low energy particle spectra will emerge in 
four dimensions as was worked out in the F-theory revival starting 
with the papers of~\cite{Donagi:2008ca,Beasley:2008dc,Beasley:2008kw,Donagi:2008kj}. 

An additional nice feature of F-theory is a largely unified 
description of gauge- or brane moduli in terms of the complex 
structure moduli space of the fourfold. Together with mirror 
symmetry this results in a large variety of geometrical tools 
that can be used to study the physically relevant structures on 
these moduli spaces. In this paper we want to improve on these tools following the line of the 
papers~\cite{Greene:1993vm,Mayr:1996sh,Klemm:1996ts,Grimm:2009ef,Alim:2009bx,Bizet:2014uua}. 

Of particular interest when studying the F-theory effective action associated to a given Calabi-Yau fourfold are the admissible fluxes.
There are two different types, namely horizontal and vertical fluxes, and in general both are necessary to construct phenomenologically viable models.
While determining a basis of fluxes over $\mathbb{C}$ is relatively straightforward, it has been shown that the fluxes are quantized~\cite{Witten:1996md} and finding 
the proper sublattice - in particular for the horizontal part - is more involved.
However, horizontal fluxes on a Calabi-Yau fourfold $W$ can be identified with the charges of topological B-branes on a mirror manifold $M$.
In this work we use the derived category description of the latter and the asymptotic charge formula in terms of the Gamma class~\cite{MR2483750,MR2683208,Kontsevichgamma} to
determine properly quantized fluxes on $W$.
We provide formulas that allow to write down the integral fluxes - and in many cases an integral basis - in terms of the intersection data on $M$.

We then restrict to the case of non-singular elliptic Calabi-Yau fourfolds and find explicit expressions for several elements of the monodromy group $\Gamma_M$.
We show that a generic subgroup of the monodromy generates the ${\rm SL}(2,\mathbb{Z})$ action on the K\"ahler modulus of the fiber.
This explains certain modular properties of the topological string amplitudes on $M$ that we also analyze in detail.
We find that the genus zero amplitudes in the type II language that determine the K\"ahler 
potential and the superpotential are ${\rm SL}(2,\mathbb{Z})$ quasi-modular forms, extending results of~\cite{Haghighat:2015qdq}.
We also show that similar features hold for the genus one amplitude, which is conjectured to be related to the gauge 
kinetic terms.
As in the Calabi-Yau threefold case we find that these amplitudes 
are related via certain holomorphic anomaly equations, from which they 
can be reconstructed in simple situations~\cite{Klemm:2004km,Klemm:2012sx,Scheidegger:2012ts,Huang:2015sta,Gu:2017ccq}.

Finally, we study the global structure of the properly quantized horizontal flux superpotential for a particular example.
To this end we analytically continue the integral periods to the generic conifold locus, the generic orbifold and the Gepner point.
We find that aligned flux stabilizes the theory at the conifold where the scalar potential vanishes.
Somewhat surprisingly the complex $8\times 8$ continuation matrix can be expressed analytically up to five real constants. 

In the rest of the introduction we describe 
the principle structures associated to the moduli space of 
Calabi-Yau fourfolds. This will set our notation and guide
the reader in later chapters, where we add to this discussion. 

\subsection{Mathematical and physical structures on the moduli space}

Let us give a very short account of the complex structure moduli 
space of Calabi-Yau fourfolds $W$, its algebraic and differential 
structures and their physical interpretation. 

As far as the differential structure and some aspects of mirror 
symmetry are concerned this is based on the analysis 
of~\cite{Greene:1993vm,Mayr:1996sh,Klemm:1996ts}. 
The analysis can be viewed as a generalization 
of the ones that lead to special geometry for Calabi-Yau 
threefolds~\cite{BG} and was discussed with emphasis on 
mirror symmetry in~\cite{Candelas:1990pi}.

Calabi-Yau manifolds are equipped with a K\"ahler $(1,1)$ form 
$\omega$ and a no-where vanishing holomophic $(4,0)$ 
form $\Omega$ with the relation $\omega^4/12=\Omega\wedge 
\bar \Omega$. The complex structure moduli space ${\cal M}$ is unobstructed
and of complex dimension $h_{3,1}(W)$. Further key structures are
the bilinear intersection form on the horizontal cohomology $\alpha_{pq}, \beta_{rs} 
\in H^4_{hor}(W)=H^{40}\oplus H^{31}\oplus H^{22}_{hor}\oplus H^{13}\oplus H^{04}$
\be 
\langle \alpha_{pq} , \beta_{rs} \rangle = \int_W \alpha_{pq} \wedge 
\beta_{rs}=0 \qquad {\rm unless} \quad p=s\ \ {\rm and} \ q=r\,,
\label{bilinear} 
\ee
which is even as the dimension is even and transversal
with respect to the Hodge type as indicated.

Moreover there is 
a positive real structure
\be
R(\alpha)= i^{p-q}\langle \alpha , \bar \alpha \rangle >0 \ ,
\label{real} 
\ee 
where $\alpha$ is a primitive form in $H^{p,q}$ with $p+q=n$. 
In particular 
\be
e^{-K(z)}= R(\Omega(z))\,,
\label{kahlerpotential} 
\ee 
defines the real K\"ahler potential $K$ for the Weil-Petersson
metric $G_{i\bar \jmath}=\partial_j \bar \partial_{\bar \jmath}K$, 
which is closely related to kinetic terms of the moduli fields
in the $N=1$ 4d effective action.
Here $\partial_j=\frac{\partial}{\partial z^i}$ or $\bar \partial_{\bar \jmath}$
are the derivatives with respect to the generic coordinates $z^i$ on ${\cal M}$ and their complex 
conjugates.

Because the intersection (\ref{bilinear}) is even on fourfolds 
one gets a mixture of algebraic and differential conditions on the periods and 
if we consider the cohomology over $\mathbb{Z}$ we get lattice structures somewhat 
similar to that of K3 surfaces.
In particular the relations
\begin{align}
\begin{split}
\int\limits_W\Omega\wedge\Omega=0\,,\quad\int\limits_W\Omega\wedge\partial_{i_1}\dots\partial_{i_n}\Omega=0\,,\quad\text{for}\quad n\le 3\,,
\label{eqn:algconstraints}
\end{split}
\end{align}
lead to non-trivial constraints on the periods.
In~\cite{Bizet:2014uua} these relations have been used to fix an integral basis for particular one parameter Calabi-Yau fourfolds.
Moreover, the authors used the Gamma class formula for the 8-brane charge as a non-trivial check of their results.
We verified that the algebraic constraints can be used to fix an integral basis for the mirror of the two-parameter elliptic Calabi-Yau fourfold $X_{24}$ but found that
this method quickly becomes unpractical if the number of moduli increases.
Our approach is somewhat complementary in that we use the Gamma class formula to fix integral periods
and the constraints \eqref{eqn:algconstraints} can be used to supplement our technique and as a non-trivial check.
In particular, this approach scales well with the number of moduli.

Other immediate data are the 4-point couplings
\be
C_{ijkl}(z) = \langle \Omega, \partial_i \partial_j \partial_k \partial_l 
\Omega(z) \rangle \ .
\label{fourpoint} 
\ee 
By the usual relation of the horizontal and vertical cohomology 
rings of $W$ to the (chiral,chiral) and (chiral,anti-chiral) 
rings of the $N=(2,2)$ superconformal theory on the 
worldsheet - with their $U(1)_l\times U(1)_r$ charge 
bigrading corresponding to the Hodge type grading~\footnote{The exchange of this identification
is the essence of mirror symmetry between $W$ and $M$. } - and the axioms of the CFT one sees however 
that these 4-point couplings are not fundamental, but 
factorize into three-point couplings
\be 
C_{ijkl}(z)=C^{\alpha}_{ij}(z) \hat{\eta}^{(2)}_{\alpha\beta} C^{\beta}_{\ kl}(z) = C^{\alpha}_{ij}(z)C^{p}_{\alpha k}(z) \hat{\eta}^{(1)}_{pl}\,,
\label{triplecouplings}
\ee
with the independent associativity condition
\be 
C^{\alpha}_{ij}(z) \hat{\eta}^{(2)}_{\alpha\beta} C^{\beta}_{kl}(z) = C^{\alpha}_{ik}(z) \hat{\eta}^{(2)}_{\alpha\beta} C^{\beta}_{jl}(z) \ . 
\label{associativity} 
\ee
Here the latin indices run over the moduli fields associated 
to either the complex structure moduli on $W$ whose tangent space is associated 
to harmonic forms in $H^{3,1}(W)$ (dual to $H^{1,3}(W)$) or K\"ahler 
moduli on $M$ whose tangent space is associated to harmonic forms in 
$H^{1,1}(M)$ (dual to $H^{3,3}(M)$). The greek indices are associated to elements in
$H_{hor}^{2,2}(W)$ and $H_{vert}^{2,2}(M)$, respectively. The $\hat{\eta}$'s define 
a constant intersection form with respect to a fixed basis of $H^{hor}_4(W)$ 
or a suitable K-theory basis extending $H^{vert}_{*,*}(M)$.

More specifically we can identify $\hat{\eta}^{(2)}$ in a reference complex structure near large radius 
with the inverse of the pairing on $H_{hor}^{2,2}(W)$ and $\hat{\eta}^{(1)}$ with the inverse pairing on 
$H^{3,1}(W)\oplus H^{1,3}(W)$, which by (\ref{bilinear}) is block 
diagonal. This property is maintained throughout the moduli space 
due to the charge grading.

The basic idea of mirror symmetry is to calculate these couplings, 
which are nontrivial sections of tensor bundles over ${\cal M}$, from 
the periods of $\Omega$. The latter can be obtained as the solutions of the Picard-Fuchs differential 
equations.
We denote an integral basis of periods by $\Pi_\kappa(z)=\int_{\Gamma^\kappa}\Omega$, where 
$\kappa=1, \ldots,{\rm dim}\, H_{hor}^4$ and ${\Gamma^\kappa}$ is a fixed 4-cycle 
basis in $H^{hor}_4(W,\mathbb{Z})$.
This is physically relevant as the flux superpotential 
\be 
W(z) =\int_W G_4 \wedge \Omega(z) = n^\kappa \Pi_\kappa(z) \,,
\label{fluxpotential}
\ee
is given with respect to this basis by (half)\footnote{As pointed out in \cite{Witten:1996md} the combination $\left[G_4 - \frac{c_2(M)}{2}\right]\in H_4(M,\mathbb{Z})$ has to be integral. 
However, in the concrete examples discussed below $c_2(M)$ is even.} integer flux quanta $n^\kappa\in \mathbb{Z}$, quantized due 
to a Dirac-Zwanziger quantization condition and additional constraints discussed 
in~\cite{Witten:1996md}. The analysis of attractor points and cosmologically
suitable minima of the associated scalar potential relies therefore crucially on this 
basis.

Interpreted in the A-model the triple couplings $C_{ij}^{\alpha}(t)$ 
in the flat coordinates given by the mirror map $t_k(z)\propto 
\int_{[C_k]} (\omega+ iB)$, where $[C_k]$ is an integral curve class on 
$M$ and $B$ is the Neveu-Schwarz B-field, encode the quantum 
cohomology of $M$.
In particular each coefficient of the Fourier expansion 
$C_{ij}^{\alpha}(e^{2 \pi i t_k})$ counts the contribution of a holomorphic 
worldsheet instanton in a given topological class. These contributions 
are directly related to Gromov-Witten invariants at genus zero.
Gromov-Witten invariants at genus one can be calculated from the 
Ray-Singer Torsion, starting with the genus zero data. Both genus 
zero and genus one worldsheet instanton series give rise to a 
remarkable integrality structure in terms of additional geometric 
invariants of embedded curves~\cite{Klemm:2007in}.

An interesting aspect of these generating functions is that they 
are modular forms of the monodromy group $\Gamma$ preserving the 
intersection form in the integral basis. For generic Calabi-Yau 
fourfolds this aspect is too difficult to appreciate in the sense that not much is known about the corresponding automorphic forms, but for 
elliptically fibered Calabi-Yau spaces, there is a
subgroup of $\Gamma$ which acts as the modular group
on the K\"ahler modulus $\tau$ of the elliptic fiber in $M$. The 
precise way this subgroup is embedded in $\Gamma$
can be inferred using specific auto-equivalences 
of the derived category of $B$-branes, as we will see in section~\ref{FourierMukai}.

It turns out that there is a clash between holomorphicity and modularity in the $\tau$ dependence of the triple couplings and the Ray-Singer torsion, which 
leads for Calabi-Yau threefolds to the holomorphic anomaly equations. We 
will discuss analogous holomorphic anomaly equations for fourfolds in section~\ref{Amplitudes}.

\section{The period geometry of Calabi-Yau fourfolds}
\label{Periodgeometry}

In this section we show how to determine integral horizontal fluxes on a Calabi-Yau fourfold $W$.
To this end we interpret the flux lattice as the charge lattice of A-branes on $W$.
This in turn is related via homological mirror symmetry to the charge lattice of B-branes on a mirror manifold $M$.
B-branes on $M$ form the bounded derived category of coherent sheaves $D^b(M)$. 
Given a brane $\mathcal{E}^\bullet\in D^b(M)$ the asymptotic behaviour of the charge can be calculated  
using the $\Gamma$-class.
Moreover, an $\mathbb{C}$-basis of fluxes on $W$ can be obtained as the solution to a set of differential equations, the Picard-Fuchs system.
Integral generators are then linear combinations of solutions with the correct asymptotic behaviour.

A similar calculation has been used in \cite{Gerhardus:2016iot} to obtain the quantum corrected A-model cohomology ring
for certain non-complete intersection Calabi-Yau fourfolds.
In some cases the asymptotic behaviour was not sufficient to uniquely determine integral elements.
As was pointed out in \cite{Gerhardus:2016iot}, the Jurkiewicz-Danilov theorem and the Lefschetz hyperplane theorem prevent this behaviour
for the induced cohomology on complete intersections in toric ambient spaces.
In general algebraic constraints on the periods can be used to supplement the above procedure. 

\subsection{The structure of $H^{4}(W,\mathbb{Z})$}
The structure of $H^4(W,\mathbb{Z}$) for a Calabi-Yau fourfold is surprisingly subtle
and in this paper we will only be interested in finding an integral basis for the period lattice.
However, even this notion demands justification.

We first discuss the structure of $H^4(W,\mathbb{C})$.
By the definition of a Calabi-Yau manifold, $H^{4,0}(W,\mathbb{C})$ is generated by a unique, holomorphic 4-form that we call $\Omega$.
Then  $H^{3,1}(W,\mathbb{C})$ is generated by first-order derivatives $\partial_{z_i}\Omega$ - modulo a part in $H^{4,0}(W,\mathbb{C})$ - where $z_i$ 
are complex structure coordinates. Due to the existence of the the harmonic $(4,0)$ form, $H^{3,1}(W,\mathbb{C})$ can be identified with the first 
order deformations of the complex structures and by the Tian-Todorov theorem  the latter are unobstructed.   
$H^{1,3}(W,\mathbb{C})$ and $H^{0,4}(W,\mathbb{C})$ are obtained from these spaces by complex conjugation.

The interesting part is thus $H^{2,2}(W,\mathbb{C})$. By Lefschetz decomposition the cohomology splits into
\begin{align}
H^{2,2}(W,\mathbb{C})=H^{2,2}_{\text{prim}}(W,\mathbb{C})\oplus H^{2,2}_V(W,\mathbb{C})\,.
\end{align}
Here the subgroup of primitive classes is given by
\begin{align}
H^{2,2}_{\text{prim}}(W,\mathbb{C})=\{\alpha\in H^{2,2}(W,\mathbb{C})\,|\,\omega\wedge\alpha =0\}\,,
\end{align}
where $\omega$ is the K\"ahler form.
On the other hand the so-called \textit{primary vertical cohomology} is generated by the ${\rm SL}(2,\mathbb{Z})$ Lefschetz action from the primitive classes in $H^{1,1}(W,\mathbb{C})$, i.e.
\begin{align}
H^{2,2}_V(W,\mathbb{C})=\{\omega\wedge\beta\,|\,\beta\in H^{1,1}(W,\mathbb{C})\,,\,\omega^3\wedge\beta=0\}\,.
\end{align}
We now denote the subspace of cohomology generated by derivatives $\partial_{z_{i_1}}\dotsm\partial_{z_{i_n}}\Omega$ of the holomorphic $4$-form as the \textit{primary horizontal cohomology} $H_H^4(W,\mathbb{C})$.
Since the K\"ahler class is independent of the complex structure, it follows from
\begin{align}
\omega\wedge\Omega=0\,,
\end{align}
that $H^{2,2}_H(W,\mathbb{C})=H^4_H(W,\mathbb{C})\cap H^{2,2}(W,\mathbb{C})$ lies inside $H_{\text{prim}}^{2,2}(W,\mathbb{C})$.
However, as was shown in \cite{Watari:2014}, there can be additional primitive classes in $H_{\text{prim}}^{2,2}(W,\mathbb{C})\backslash H_H^{2,2}(W,\mathbb{C})$.
The structure is thus
\begin{align}
H^{2,2}(W,\mathbb{C})=H_H^{2,2}(W,\mathbb{C})\oplus H_{RM}^{2,2}(W,\mathbb{C})\oplus H_V^{2,2}(W,\mathbb{C})\,,
\end{align}
where $H_{RM}^{2,2}(W,\mathbb{C})$ is the subgroup of primitive classes that are neither horizontal nor vertical.

The naive expectation that mirror symmetry maps vertical into horizontal classes and vice versa
while the remaining component maps into itself can not hold.
It would lead to a contradiction when applied to the geometry studied in \cite{Gerhardus:2016iot}, where additional ``vertical'' cycles appear in the quantum deformed A-model intersections.
A true statement about the relation under mirror symmetry would therefore require a more refined notion of verticality.
This subtlety is avoided when phrasing the problem in terms of branes and homological mirror symmetry.

\subsection{Fixing an integral basis}
\label{Fixingtheintegralbasis}
A $4$-cycle $\Sigma$ dual to an element in $H^4_H(W,\mathbb{C})\cap H^4(W,\mathbb{Z})$ is calibrated symplectically, i.e.
\begin{align}
\left.\text{Re}\,e^{i\theta}\Omega\right|_\Sigma=0\,,
\end{align}
and the K\"ahler class restricts to zero $\omega|_\Sigma=0$. In other words, $\Sigma$ is a special lagrangian cycle that can be wrapped by a topological A-brane $L$.
The central charge of this brane is then given by the period
\begin{align}
Z_A(L)=\int\limits_\Sigma\Omega\,.
\end{align}
Note that this is equal to the superpotential generated by a flux quantum along $\Sigma$.

By homological mirror symmetry~\cite{MR1403918,Strominger:1996}, the topological A-branes on $W$ are related to B-branes on the mirror $M$.
The latter correspond to elements in $D^b(M)$, the bounded derived category of coherent sheaves on $M$.
Given a B-brane that corresponds to a complex $\mathcal{E}^\bullet\in D^b(M)$, the asymptotic behaviour of the central charge is
\begin{align}
Z_B^{\text{asy}}(\mathcal{E}^\bullet)=\int_M e^J\Gamma_\mathcal{\mathbb{C}}(M)\,({\text{ch}\,\mathcal{E}^\bullet})^\lor\,,
\end{align}
where $J$ is the K\"ahler class on $M$. The details of this formula will be discussed in the next section.
The crucial fact is that the central charges of A- and B-branes are identified via the mirror map.
While a construction for all objects in $D^b(M)$ is in general not available, the central charge only depends on the K-theory charge of a complex of sheaves.

Our approach to fix an integral basis for the period lattice will be to construct elements $\mathcal{E}^\bullet$ in $D^b(M)$ that generate the algebraic K-theory group $K^0_\text{alg}(M)$ and 
calculate the asymptotic behaviour of the central charges.
Using the mirror map, these can be interpreted as the leading logarithmic terms of generators of the period lattice.
The subleading terms are given by the corresponding solutions to the Picard-Fuchs equations.
\subsection{B-branes and the asymptotic behaviour of the central charge}
\label{Gammaclass} 
For a Calabi-Yau manifold $M$, the topological B-branes and the open string states stretched between them are encoded in the bounded derived category of coherent sheaves $D^b(M)$.
The objects of this category are equivalence classes of bounded complexes of coherent sheaves
\begin{align}
\xymatrix{
\mathcal{E}^\bullet=\ldots\ar[r]^{d^{\mathcal{E}}_{-2}}&\mathcal{E}^{-1}\ar[r]^{d^{\mathcal{E}}_{-1}}&\mathcal{E}^0\ar[r]^{d^{\mathcal{E}}_{0}}&\mathcal{E}^{1}\ar[r]^{d^{\mathcal{E}}_{1}}&\ldots\,.
}
\end{align}
A set of maps $f_i:\mathcal{E}^i\rightarrow\mathcal{F}^i$, such that the $f_i$ commute with the coboundary maps, corresponds to an element $f\in\text{Hom}(\mathcal{E}^\bullet,\mathcal{F}^\bullet)$.
Objects as well as morphisms are identified under certain equivalence relations but a more detailed discussion of topological branes and $D^b(M)$ is outside the scope of this paper and can be found e.g. in \cite{Aspinwall:2004jr}.

However, we note that if there is an exact sequence
\begin{align} 
\xymatrix{
\ldots\ar[r]&\mathcal{E}^{-1}\ar[r]&\mathcal{E}^0\ar[r]&\mathcal{F}\ar[r]&0\,,
}
\end{align}
where $\mathcal{F}$ is a coherent sheaf and $\mathcal{E}^i$ are locally free sheaves, i.e. equivalent to vector bundles, then the complex
\begin{align} 
\xymatrix{
\mathcal{E}^\bullet=\ldots\ar[r]&\mathcal{E}^{-1}\ar[r]&\mathcal{E}^0\ar[r]&0\,,
}
\end{align}
is equivalent to $\mathcal{F}$ inside $D^b(M)$.

Now given the K\"ahler class $J$, the asymptotic charge of a B-brane that corresponds to the complex $\mathcal{E}^\bullet$ is given by
\begin{align}
Z^\text{asy}({\mathcal{E}^\bullet})=\int_M e^J\Gamma_\mathbb{C}(M)\left(\text{ch}\,\mathcal{E}^{\bullet}\right)^\vee\,.
\end{align}
The characteristic class $\Gamma_\mathbb{C}(M)$ can be expressed in terms of the Chern classes of $M$ and for a Calabi-Yau manifold the expansion reads
\begin{align}
\Gamma_\mathbb{C}(M)=1+\frac{1}{24}c_2-\frac{i\zeta(3)}{8\pi^3}c_3+\frac{1}{5760}(7c_2^2-4c_4)+\dots\,.
\end{align}
The Chern character of the complex is given by
\begin{align}
\text{ch}(\mathcal{E}^\bullet)=...-\text{ch}(E^{-1})+\text{ch}(E^0)-\text{ch}(E^1)+\text{ch}(E^2)-\dots\,,
\end{align}
where $E^i$ is the vector bundle corresponding to the locally free sheaf $\mathcal{E}^i$
and the involution $(\dots)^\vee$ acts on an element $\beta\in H^{2k}(M)$ as $\beta^\vee=(-1)^k\beta$.

A general basis of $0$-, $2$-, $6$- and $8$-branes has been constructed in~\cite{Gerhardus:2016iot}.
The $8$-brane corresponds to the structure sheaf $\mathcal{O}_M$ and the $6$-branes are generated by locally free resolutions of sheaves $\mathcal{O}_{J_i}$, where the divisors $J_i$ generate the K\"ahler cone.
The $0$-brane is represented by the skyscraper sheaf $\mathcal{O}_{\text{pt.}}$.
A basis of $2$-branes was constructed as
\begin{align}
\mathcal{C}^\bullet_a = \iota_! \mathcal{O}_{\mathcal{C}^a}(K_{\mathcal{C}^a}^{1/2}) \ ,
\end{align}
where $\iota$ is the inclusion of the curve $\mathcal{C}^a$ that is part of a basis for the Mori cone and $K_{\mathcal{C}^a}^{1/2}$ is a spin structure on $\mathcal{C}^a$.
The asymptotic charges have been calculated in \cite{Gerhardus:2016iot} and for the readers convenience they are reproduced below.

We now describe a construction of $4$-branes which in many cases leads to an integral basis.
Given effective divisors $D_i,\,i\in I$ that correspond to codimension one subvarieties of $M$ and $S=\bigcap\limits_{i\in I}D_i$, the Koszul sequence
\begin{align}
\parbox{2cm}{
\xymatrix{
&0\ar[r]&\smash{\mathcal{O}_M\left(-\sum\limits_{\mclap{i\in I}}D_i\right)}\ar[r]&\smash{\underset{j\in I}{\oplus}\mathcal{O}_M\left(-\sum\limits_{\mclap{i\in I\backslash \{j\}}}D_i\right)}\ar[r]&\ldots\ar`[rd]`[l]`[dllll]`[d][dlll]\\
&\smash{\underset{i\in I}{\oplus}\mathcal{O}_M\left(-D_i\right)}\ar[r]&\smash{\mathcal{O}_M}\ar[r]&\smash{\mathcal{O}_{S}}\ar[r]&0&
}
}
\label{eqn:koszul}
\end{align}
is exact and provides a locally free resolution of the coherent sheaf $\mathcal{O}_S$.
When $I$ contains only one element, this is just the familiar short exact sequence
\begin{align}
\xymatrix{
0\ar[r]&\mathcal{O}_M(-D)\ar[r]&\mathcal{O}_M\ar[r]&\mathcal{O}_D\ar[r]&0\,.
}
\end{align}
The latter implies the equivalence
\begin{align}
\xymatrix{
0\ar[r]&\mathcal{O}_M(-D)\ar[r]&\mathcal{O}_M\ar[r]&0\quad\sim\quad0\ar[r]&\mathcal{O}_D\ar[r]&0\,,
}
\end{align}
of complexes in $D^b(M)$.
This is the locally free resolution employed in \cite{Gerhardus:2016iot} to calculate the central charges for a basis of $6$-branes.

More generally, we can use the Koszul sequence to describe branes wrapped on arbitrary cycles that are intersections of subvarieties of codimension one.
If a basis of $H^{2,2}_V(M,\mathbb{C})\cap H^4(M,\mathbb{Z})$ can be constructed this way, then, as we described above, this leads to an integral basis of the period lattice in the mirror.
In particular the asymptotic behaviour then uniquely singles out a solution to the Picard-Fuchs system.
For a Calabi-Yau hypersurface $M$ in a toric variety $\mathbb{P}_\Delta$, the cohomology of the ambient spaces is generated by elements in $H^{1,1}(\mathbb{P}_\Delta)$.
As was pointed out by the authors of \cite{Gerhardus:2016iot}, the quantum Lefschetz hyperplane theorem then guarantees that $H_V^{2,2}(M,\mathbb{C})$ is
generated by restrictions of elements in $H^{2,2}(\mathbb{P}_\Delta,\mathbb{C})$.

The formula for the asymptotic central charge gives the following results:
\begin{itemize}
\item \textbf{8-brane:}
\begin{align}
\begin{split}
Z^{\text{asy}}\left(\mathcal{O}_M\right)=&\int_M e^J\Gamma_\mathbb{C}(M)=\frac{1}{4!}C_{ijkl}^0t^it^jt^kt^l+\frac{1}{2}c_{ij}t^it^j+c_it^i+c_0\,,\\
C_{ijkl}^0=&\int_M J_iJ_jJ_kJ_l\,,\quad c_{ij}=\frac{1}{24}\int_Mc_2(M)J_iJ_j\,,\\
c_i=&-\frac{i\zeta(3)}{8\pi^3}\int_Mc_3(M)J_i\,,\quad c_0=\frac{1}{5760}\int_M\left[7c_2(M)^2-4c_4(M)\right]
\end{split}
\end{align}
\item \textbf{6-brane wrapped on $J_a$:}
{\small
\begin{align}
\begin{split}
Z^{\text{asy}}\left(\mathcal{O}_{J_a}\right)=&\int_M e^J\Gamma_\mathbb{C}(M)\left[1-\text{ch}\left(\mathcal{O}_M(J_a)\right)\right]\\
=&-\frac{1}{3!}C^0_{aijk}t^it^jt^k-\frac{1}{4}C^0_{aaij}t^it^j-\left(\frac{1}{6}C^0_{aaai}+\frac{1}{24}c^a_i\right)t^i\\
&-\left(\frac{1}{24}C^0_{aaaa}+c^a_0\right)\,,\\
c^a_i=&\int_Mc_2(M)J_a J_i\,,\quad c^a_0=\frac{1}{48}\int_M c_2(M)J_a^2-\frac{\zeta(3)}{(2\pi i)^3}\int_M c_3(M)J_a
\end{split}
\end{align}
}
\item \textbf{4-brane wrapped on $H=D_a\cap D_b$:}

\begin{align}\label{4-asy}
\begin{split}
Z^{\text{asy}}\left(\mathcal{O}_{D_a\cap D_b}\right)=&\frac{1}{2}\int_Mh_{ij}t^it^j+h_it^i+h\,,\\
h_{ij}=&\int_MD_aD_bJ_iJ_j\,,\quad h_i=\frac12\int_MD_aD_b(D_a+D_b)J_i\,,\\
h=&\frac{1}{12}\int_MD_aD_b(2D_a^2+3D_aD_b+2D_b^2)+\frac{1}{24}\int_Mc_2(M)D_aD_b
\end{split}
\end{align}

\item \textbf{2-brane wrapped on $\mathcal{C}^a$ dual to $J_a$:} 
\begin{align}
Z_{\text{asy}}(\mathcal{C}_a^\bullet)=-t_a
\end{align}
\end{itemize}
The charge of the \textbf{0-brane} is universally $Z^{\text{asy}}(\mathcal{O}_\text{pt.})=1$.
We denoted the generators of the K\"ahler cone by $J_i$ and the K\"ahler form is given by $J=t^iJ_i$.

Finally we need the intersection matrix of the $4$-cycles mirror dual to the B-branes.
They are not given by the classical intersection numbers in the A-model but rather by the
open string index
\begin{align}
\chi(\mathcal{E}^\bullet,\mathcal{F}^\bullet)=\int_M\text{Td}(M)\left(\text{ch}\,\mathcal{E}^{\bullet}\right)^\vee\text{ch}\,\mathcal{F}^\bullet\,.
\end{align}
The Todd class $\text{Td}(M)$ is for a Calabi-Yau fourfold given by
\begin{align}
\text{Td}(M)=1+\frac{c_2(M)}{12}+2V\,,
\end{align}
where $V$ is the volume form.
Note that if we construct a basis of B-branes
\begin{align}
\vec{v}=(\mathcal{E}_1^\bullet,\dots,\,\mathcal{E}_n^\bullet),
\end{align}
and introduce the intersection matrix $\eta_{ij}=\chi(v_i,v_j)$, the inverse matrix $\eta^{-1}$ will act on
the period vector $\Pi$ corresponding to the mirror dual cycles.
For example
\begin{align}
\int_W\Omega\wedge\Omega=0\quad\rightarrow\quad\Pi^T\eta^{-1}\Pi=0\,.
\end{align} 

\section{Elliptically fibered Calabi-Yau fourfolds}
\label{EllipticallyfibredCalabiYau}
Although the methods to find integral generators of the period lattice are applicable to general Calabi-Yau manifolds
we now restrict to elliptic fibrations
\begin{align}
\xymatrix{
\mathcal{E}\ar[r]&M\ar[d]^{\pi}\\
&B
}
\end{align}
such that for a general choice of complex structure on $M$ the fiber exhibits at most $I_1$ singularities over loci of codimension $1$ in the base $B$.
In particular we require the presence of a section.
Fourfolds of this type have been previously studied in \cite{Haghighat:2015qdq}.
It turns out that the intersection ring and the relevant topological invariants are completely determined by the base.
Note that this geometric setup is completely analogous to the threefolds studied in \cite{Klemm:2012sx,Huang:2015sta}.

\subsection{Geometry of non-singular elliptic Calabi-Yau fourfolds}
\label{BasicPropertiesellipticfibre} 
As far as it carries over to the fourfold case, we follow the notation in \cite{Huang:2015sta} which we now quickly review.
The generators of the Mori cone of the base $B$ are given by $\{[\tilde{\mathcal{C}}'^k]\},\,k=1,\ldots,h_{11}(B)=h_{11}(M)-1$ and the dual basis of the K\"ahler cone is $\{[D_k']\}$.
In particular we assume that the Mori cone is simplicial.
Let $E$ be the section so that its divisor class is given by $[E]$.

We now obtain curves
\begin{align}
\tilde{\mathcal{C}}^k=E\cdot\pi^{-1}\tilde{\mathcal{C}}'^k\,,\,k=1,\ldots,h_{11}(B)\,,
\end{align}
on $M$ for some representatives $\tilde{\mathcal{C}}'^k$ of $[\tilde{\mathcal{C}}'^k]$.
A basis for the Mori cone on $M$ is given by $\{[\tilde{\mathcal{C}}^k],[\tilde{\mathcal{C}}^e]\}$, where $[\tilde{\mathcal{C}}^e]$ is the class of the generic fiber.
The K\"ahler cone of $M$ is generated by the dual basis $\{[\tilde{D}_e],[\tilde{D}_k]\}$, where
\begin{align}
[\tilde{D}_k]=\pi^*[D_k']\,,\quad[\tilde{D}_e]=[E]+\pi^*c_1(B)\,.
\end{align}
In the following we will mostly drop the square brackets and assume that the distinction between subvarieties and corresponding classes is clear from the context. 
The intersection ring of $M$ is determined in terms of intersections on $B$ via
{\small
\begin{align}
\begin{split}
\int_M \tilde{D}_e\cdot P(\tilde{D}_e,\tilde{D}_1,\ldots,\tilde{D}_{h_{11}(B)})=&\int_B P(c_1(B),D_1',\ldots,D_{h_{11}(B)}')\,,\\
\int_M P(1,\tilde{D}_1,\ldots,\tilde{D}_{h_{11}(B)})=&\,0\,,
\end{split}
\end{align}
}
where $P$ is any polynomial in $h_{11}(B)+1$ variables.

We denote the complexified areas of the curves in the base by
\begin{align}
\tilde{T}^k=\int_{\tilde{\mathcal{C}}^k}\mathcal{B}+i\omega\,,
\end{align}
where $\omega$ is the K\"ahler class and $\mathcal{B}$ is the Neveu-Schwarz $\mathcal{B}$-field.
The complexified area of the fiber will be called
\begin{align}
\tilde{\tau}=\int_{\mathcal{\tilde{C}}^e}\mathcal{B}+i\omega\,.
\end{align}

The generators of the Mori cone and the dual generators of the K\"ahler cone provide a natural choice of basis for divisors and curves from the geometric perspective.
However, as was already observed for elliptically fibered threefolds, the ${\rm SL}(2,\mathbb{Z})$ subgroup of the monodromy acts more naturally in a different choice of basis.
We introduce
\begin{align}
[\mathcal{C}^e]=[\tilde{\mathcal{C}}^e]\,,\quad[\mathcal{C}^k]=[\tilde{\mathcal{C}}^k]+\frac{a^k}{2}[\tilde{\mathcal{C}}^e]\,,
\end{align}
with
\begin{align}
a^k=\int_{\tilde{\mathcal{C}}^k}c_1(B)\,,
\end{align}
and the dual basis
\begin{align}
D_e=\tilde{D}_e-\frac12\pi^*c_1(B)=E+\frac12\pi^*c_1(B)\,,\quad D_k=\tilde{D}_k\,.
\end{align}
The complexified areas corresponding to $\mathcal{C}^k$ and $\mathcal{C}^e$ are now given by
\begin{align}\label{spcoord}
\tau=\tilde{\tau}\quad\text{and}\quad T^k=\tilde{T}^k+\frac{a^k}{2}\tilde{\tau}\,,
\end{align}
respectively.
Finally we introduce the exponentiated complexified areas
\begin{align}
\tilde{Q}^k=\exp(2\pi i \tilde{T}^k)\,,\quad \tilde{q}_e=\exp(2\pi i\tilde{\tau})\,,
\end{align}
with similar definitions for $Q^k$ and $q_e$.

We also define the topological invariants of the base
\begin{align}\label{constants} 
\begin{split}
a=c_1(B)^3\,,\quad a_i=c_1(B)^2\cdot D_i'\,,\quad a_{ij}=c_1(B)\cdot D_i'\cdot D_j'\,,\quad c_{ijk}=D_i'\cdot D_j'\cdot D_k'\,,
\end{split}
\end{align}
and denote the $k$-th degree component of $\text{ch}(\mathcal{F}^\bullet)$ by $\text{ch}_k(\mathcal{F}^\bullet)$.

The definitions above are straightforward extensions of the corresponding threefold expressions introduced in \cite{Huang:2015sta}.
For Calabi-Yau fourfolds a basis of middle-dimensional cycles has to be specified as well. It turns out that for elliptically fibered fourfolds with at most $I_1$ singularities in the fibers
such a basis is given by
\begin{align}\label{4cyclesbasis}
H_k=E\cdot \pi^{-1}D_k'=E\cdot\tilde{D}_k\,,\quad H^k=\pi^{-1}\tilde{C}'^k\,,
\end{align}
with
\begin{align}\label{4cyclesduality}
H_i\cdot H_j= -a_{ij}\,,\quad H_i\cdot H^j=\delta_i^j\,,\quad H^i\cdot H^j=0\,.
\end{align}
We call the 4-cycles $H^k=\pi^{-1} \tilde C^{\prime k}$, $k=1,\dots, h_{11}(B)$  that result  from lifting a 
curve in the base to a 4-cycle in $M$ the  $\pi${\sl --vertical 4-cycles}. As we will 
see in section~\ref{multip} the genus zero  amplitudes that correspond to 
$\pi${--vertical 4-cycles} have particularly simple modular properties. Using the Koszul sequence \eqref{eqn:koszul} we calculate
\begin{align}
\begin{split}
\text{ch}(\mathcal{O}_{H_i})=&H_i-\frac{1}{2}\tilde{\mathcal{C}}^k(c_{kii}-a_{ki})+\frac{1}{12}V(2a_i-3a_{ii}+2c_{iii})\,,\\
\text{ch}(\mathcal{O}_{H^i})=&H^i-\tilde{\mathcal{C}}^e\cdot h^i\,,
\end{split}
\end{align}
with the volume form $V$ and
\begin{align}
h^i=\int_M E\,\text{ch}_3(\mathcal{O}_{H^i})=\sum\limits_{a,b}\frac12 \lambda_{a,b}E\cdot(D_a\cdot D_b)\cdot(D_a+D_b)\,,
\end{align}
where we assume that
\begin{align}
H^i=\sum_{a,b}\lambda_{a,b}\bar{D}_a\cdot \bar{D}_b\,,
\end{align}
for effective divisors $\bar{D}_a$. The Chern characters of the 6-branes are given by
\begin{align}
\begin{split}
\text{ch}(\mathcal{O}_{\tilde{D}_i})=&\tilde{D}_i-\frac12H^kc_{kii}+\frac16\tilde{\mathcal{C}}^ec_{iii}\,,\\
\text{ch}(\mathcal{O}_{E})=&E+\frac12 H_i\cdot a^i+\frac16\tilde{\mathcal{C}}^ia_i+\frac{1}{24}V\cdot a\,.
\end{split}
\end{align}
Moreover, $\text{ch}(\mathcal{O}_M)=1$, $\text{ch}(\tilde{\mathcal{C}}^{e\bullet})=\tilde{\mathcal{C}}^e$ and $\text{ch}(\tilde{\mathcal{C}}^{k\bullet})=\tilde{\mathcal{C}}^k$.
\subsection{Fourier-Mukai transforms and the ${\rm SL}(2,\mathbb{Z})$ monodromy}
\label{FourierMukai} 
The B-model periods are multi-valued and experience monodromies along paths encircling special divisors in the complex structure moduli space.
Homological mirror symmetry~\cite{MR1403918} implies that the corresponding monodromies in the A-model lift to auto-equivalences of the derived category~\cite{MR1403918,MR1876078,MR1831820}.
Furthermore, an important theorem by Orlov states that every equivalence of derived categories of coherent sheaves of smooth projective varieties is a Fourier-Mukai transform.

A Fourier-Mukai transform $\Phi_\mathcal{E}: D^b(X)\rightarrow D^b(Y)$ is determined by an object $\mathcal{E}\in D^b(X\times Y)$ and acts 
as~\cite{MR1876078,MR1831820}\footnote{An accessible explanation for physicists of how these calculations are performed can be found in \cite{Distler:2002ym}.}
\begin{align}
\mathcal{F}^\bullet\mapsto R\pi_{1*}(\mathcal{E}\otimes_L L\pi_2^*\mathcal{F}^\bullet)\,,
\end{align} 
where $\pi_1$ and $\pi_2$ are the projections from $X\times Y$ to $Y$ and $X$ respectively. 
The object $\mathcal{E}$ is called the kernel and $R$ and $L$ indicate that one has to take the left- or right derived functor in place of $\pi_*,\,\pi^*$ or $\otimes$.

For our purpose the nice property of this picture is that certain general monodromies correspond to generic Fourier-Mukai kernels.
This allows us to write down closed forms not only for the large complex structure monodromies but also for a certain generic conifold monodromy and a third type that is special
to elliptically fibered Calabi-Yau.

Let $D$ be one of the generators of the K\"ahler cone and $C$ the dual curve. The limit in which $C$ becomes large corresponds to a divisor in the K\"ahler moduli space.
It is well known~\cite{MR1876078} that the Fourier-Mukai transform corresponding to the monodromy around this large radius divisor acts as
\begin{align}
\mathcal{E}^\bullet\mapsto \mathcal{O}(D)\otimes\mathcal{E}^\bullet\,.
\end{align}
We choose a basis of branes
\begin{align}
\left(\mathcal{O}_M,\mathcal{O}_E,\mathcal{O}_{D_i},\mathcal{O}_{H_i},\mathcal{O}_{H^i},\tilde{\mathcal{C}}^i,\tilde{\mathcal{C}}^e,\mathcal{O}_{\text{pt.}}\right)\,,
\end{align}
and calculate the monodromy for the large radius divisor corresponding to $D_j$, 
\begin{align}
\tilde{T}_j=\left(\begin{array}{cccccccc}
1&0&-\delta^k_j&0&0&0&0&0\\
0&1&0&-\delta^k_j&0&0&0&0\\
0&0&\delta^k_i&0&-c_{jik}&0&0&0\\
0&0&0&\delta^k_i&0&-c_{jik}&0&\frac12(c_{jii}+c_{jji}-a_{ji})\\
0&0&0&0&\delta_k^i&0&-\delta_{j}^i&0\\
0&0&0&0&0&\delta_k^i&0&-\delta^i_j\\
0&0&0&0&0&0&1&0\\
0&0&0&0&0&0&0&1
\end{array}\right)\,,
\end{align}
acting on the vector of charges.
One can obtain a similar expression for the monodromy $\tilde{T}_e$, corresponding to $\tilde{D_e}$.

Another auto-equivalence, the Seidel-Thomas twist, corresponds to the locus where, given a suitable loop based on the point of large radius, the D8-brane becomes massless.
Its action on the brane charges is given by
\begin{align}
Z(\mathcal{E}^\bullet)\mapsto Z(\mathcal{E}^\bullet)-\chi(\mathcal{E}^\bullet,\mathcal{O}_M)Z(\mathcal{O}_M)\,.
\end{align}
As was explained in~\cite{Bizet:2014uua}, for a Calabi-Yau fourfold $\chi(\mathcal{O}_M,\mathcal{O}_M)=2$.
This implies that $Z(\mathcal{O}_M)$ transforms into $-Z(\mathcal{O}_M)$ and this monodromy is of order two.

Elliptically fibered Calabi-Yau manifolds with at most $I_1$ singularities exhibit yet another type of auto-equivalence. 
Physically it corresponds to T-duality along both circles of the fiber torus.
The corresponding action $\Phi$ on the derived category was first studied by Bridgeland \cite{1997alg.geom..5002B} in the context of elliptic surfaces.
Calculations for Calabi-Yau threefolds can be found in \cite{Andreas:2000sj} and were elaborated on in the subsequent review \cite{Andreas:2004uf}. 
In full generality the auto-equivalences and their implications for the modularity of the amplitudes on elliptic Calabi-Yau threefolds with $I_1$ singularities~\cite{HKK} 
have been presented in~\cite{Katz:2016SM}.

We can decompose the Chern character of a general brane $\mathcal{E}^\bullet$ as
\begin{align}
\begin{split}
\text{ch}_0(\mathcal{E}^\bullet)=&n,\\
\text{ch}_1(\mathcal{E}^\bullet)=&n_E\, E+F_1,\\
\text{ch}_2(\mathcal{E}^\bullet)=&E\cdot B_1+F_2,\\
\text{ch}_3(\mathcal{E}^\bullet)=&E\cdot B_2+n_e\,\tilde{\mathcal{C}}^e,\\
\text{ch}_4(\mathcal{E}^\bullet)=&s\,V.
\end{split}
\end{align}
Here we introduced $n,n_E,n_e,s\in\mathbb{Q}$, and $F_i, B_i$ are pullbacks of forms in $H^{i,i}(B,\mathbb{C})$.
The volume form on $M$ is denoted by $V$.
Adapting the calculation in \cite{Andreas:2004uf} to Calabi-Yau fourfolds, we find that the Chern character of the transformed brane is given by 
\begin{align}
\begin{split}
\text{ch}_0(\Phi(\mathcal{E}^\bullet))=&n_E\,,\\
\text{ch}_1(\Phi(\mathcal{E}^\bullet))=&B_1-\frac{1}{2}n_E\, c_1-n\cdot E\,,\\
\text{ch}_2(\Phi(\mathcal{E}^\bullet))=&B_2-\frac{1}{2}B_1\cdot c_1+\frac{1}{12}n_E\, c_1^2-F_1\cdot E+\frac{1}{2}n\,c_1\cdot E\,,\\
\text{ch}_3(\Phi(\mathcal{E}^\bullet))=&-\frac{1}{2}B_2\cdot c_1+\frac{1}{12}B_1\cdot c_1^2+s\,\tilde{\mathcal{C}}^e+\frac{1}{2}\,c_1\cdot F_1\cdot E-F_2\cdot E-\frac{1}{6}n\,c_1^2\cdot E\,,\\
\text{ch}_4(\Phi(\mathcal{E}^\bullet))=&-n_e\, V-\frac{1}{6}\,c_1^2\cdot F_1\cdot E+\frac{1}{2}c_1\cdot F_2\cdot E+\frac{1}{24}n\,c_1^3\cdot E\,,
\end{split}
\end{align}
with $c_1=\pi^*c_1(B)$.
Using the formulae for the Chern characters of the basis of branes introduced above, this translates into the matrix
{\small
\begin{align}
\tilde{S}=\left(\begin{array}{cccccccc}
0&-1&0&a^k&0&\frac12\left(c_{kii}a^i-a_k\right)&0&\frac{1}{12}(3a_{ii}a^i-2c_{iii}a^i-a)\\
1&0&0&0&0&0&0&0\\
0&0&0&-\delta^k_{i}&0&a_{ki}&0&-\frac12 a_{ii}\\
0&0&\delta_{i}^k&0&0&0&0&0\\
0&0&0&0&0&-\delta_{k}^i&0&h^i+\frac12 a^i\\
0&0&0&0&\delta^{i}_k&0&h^i-\frac12 a^i&0\\
0&0&0&0&0&0&0&-1\\
0&0&0&0&0&0&1&0
\end{array}\right)\,,
\label{eqn:fmellipticmatrix}
\end{align}
}
for the corresponding monodromy.

We can now explicitly calculate that
\begin{align}
\left(\prod\limits_{i=1}^{h_{11}(B)}\tilde{T}_i^{-a^i}\right)\tilde{S}\cdot \tilde{S}=-\mathbb{I}\,,
\end{align}
and another careful calculation reveals
\begin{align}
(\tilde{S}\cdot \tilde{T}_e^{-1})^3=-\mathbb{I}\,.
\end{align}
It follows that
\begin{align}\label{modgroup}
S=\left(\prod\limits_{i=1}^{h^{11}(B)}\tilde{T}_i^{-a^i/2}\right)\tilde{S}\,,\quad T=\left(\prod\limits_{i=1}^{h^{11}(B)}\tilde{T}_i^{a^i/2}\right)\tilde{T}_e^{-1}\,,
\end{align}
generate a group isomorphic to $PSL(2,\mathbb{Z})$, the modular group.
In particular, $Q^k, q$ are invariant under $T$, while some of the $\tilde{Q}^k$ obtain a sign under $T$-transformations if the canonical class of the base is not even.
As was already noted by \cite{Huang:2015sta}, this makes $Q^k$ and $q$ the correct expansion parameters for the topological string amplitudes to exhibit modular properties. 

\subsection{Toric construction of mirror pairs}
\label{Toricconstruction} 

To fix conventions we will briefly review the Batyrev construction of Calabi-Yau $n$-fold mirror pairs ($M,\,W$) as hypersurfaces in toric ambient spaces~\cite{Batyrev:1994hm}.

The data of the mirror pair is encoded in an $n+1$-dimensional reflexive lattice polytope $\Delta\subset\Gamma$
and the choice of a regular star triangulation of $\Delta$ and the polar polytope
\begin{align}
\Delta^{*} = \{p \in \Gamma^{*}_{\mathbb{R}} \,\vert\, \langle q,p \rangle \geq -1, \forall q \in \Delta \}\,,
\end{align}
that is embedded in the dual lattice $\Gamma^{*}$.
We denoted the real extensions of the lattices by $\Gamma_\mathbb{R}$ and $\Gamma^{*}_\mathbb{R}$ respectively.
The triangulation of $\Delta^{*}$ leads to a fan by taking the cones over the facets that in turn is associated to a toric variety $\mathbb{P}_{\Delta}$.
he family $M$ of Calabi-Yau $n$-folds is given by the vanishing loci of sections $P_{\Delta} \in \mathcal{O}(K_{\Delta^{*}})$\,
\begin{equation}\label{eq:CYhypersurface}
P_{\Delta} = \sum_{\nu \in \Delta \cap \Gamma} \prod_{\nu^{*} \in \Delta^{*} \cap \Gamma^{*}}a_{\nu} x_{\nu^*}^{\langle \nu,\nu^{*}\rangle +1} = 0\,.
\end{equation}
The mirror family $W$ is obtained by exchanging $\Delta \leftrightarrow \Delta^{*}$. 

Even for a generic choice of section the Calabi-Yau varieties thus constructed might be singular.
For $n\le 3$ the singularities can be resolved by blowing up the ambient space. 
This is not always possible for fourfolds.
However, all models studied in this paper can be fully resolved by toric divisors.

\subsection{Toric geometry of elliptic fibrations}
\label{Toricelliptic}

For F-theory we need Calabi-Yau manifolds that are elliptically fibered.
One way to construct these is by taking a torically fibered ambient space such that the hypersurface constraint cuts out a genus one curve from the fiber~\cite{Braun:2011ux}. 
Toric fibrations can be understood in terms of toric morphisms.
 toric morphism $\phi:\,\mathbb{P}_\Delta\rightarrow\mathbb{P}_{\Delta_B}$ in turn is encoded in a lattice morphisms
\begin{align}
\phi:\,\Gamma\rightarrow\Gamma_B\,,
\end{align}
such that the image of every cone in $\Sigma$ is completely contained inside a cone of $\Sigma_B$.
We obtain a fibration with the fan of the generic fiber given by $\Sigma_F\in\Gamma_F$ if
the morphism $\phi: \Gamma \rightarrow \Gamma_{B}$ is surjective and the sequence 
\begin{align}
0\rightarrow \Gamma_{F} \xhookrightarrow{}\Gamma \sequence{$\phi_{B}$} \Gamma_{B}\rightarrow 0\,,
\end{align}
is exact.

We can now obtain elliptically fibered mirror pairs $(M,W)$ from the following construction~\cite{Bizet:2014uua}. First we combine a base polytope $\Delta^{B}$ and a reflexive fiber polytope $\Delta^{F}$ and embed them into a $n+1$-dimensional polytope $\Delta$ as follows:
\begin{equation}\label{eq:mirrortoricfib}
 \nu^{*}\in \Delta_{}^{*} \mathlarger{\mathlarger{\mathlarger{\mathlarger{ \Biggl\{ }}}} \left|
 \begin{array}{c;{2pt/2pt}c;{0pt/0pt}r;{2pt/2pt}c}
 \mbox{ $\Delta_{}^{B*}$} & \begin{matrix} \nu_{i}^{F*} \\ \vdots\\ \nu_{i}^{F*} \end{matrix} & \mbox{ $s_{ij}\Delta_{}^{B}$} & \begin{matrix} \nu_{j}^{F}\\ \vdots \\ \nu_{j}^{F} \end{matrix}\\
 \mbox{ \Large 0}& \Delta_{}^{F*} & \mbox{ \Large 0}\qquad& \Delta_{}^{F}
\end{array}
\right|
 \mathlarger{\mathlarger{\mathlarger{\mathlarger{ \Biggl\} }}}}\nu^{}\in \Delta_{}^{}
\end{equation}
For a fixed $\nu_{i}^{F*}\in\Delta^{F*}$ and $\nu_{j}^{F} \in \Delta^{F}$ we introduced $s_{ij} = \langle \nu_{j}^{F}, \nu_{i}^{F*}\rangle + 1 \in \mathbb{Z}_{>0}$.
This describes a reflexive pair of polytopes $(\Delta,\Delta^{*})$ given by the convex hulls of the points appearing in (\ref{eq:mirrortoricfib}).
Using the Batyrev construction one gets an $n$-fold $M$ from the locus given by (\ref{eq:CYhypersurface}) on the ambient space $\mathbb{P}_{\Delta}$. As mentioned above, $M$ inherits a fibration structure from the ambient spaces $\mathbb{P}_{\Delta}\rightarrow \mathbb{P}_{\Delta^{B}}$ and we can identify a map
\begin{equation}
M = \{ \underline{x} \subset \mathbb{P}_{\Delta}\vert P_{\Delta_{}}(\underline{x})=0\} \sequence{$\pi$} B = \mathbb{P}_{\Delta^{B}}
\end{equation}

In the following we will consider fibers constructed as $E_8$ hypersurfaces
\begin{align}
 E_{8}: & \quad X_{6}(1,2,3)=\{(x,y,z) \subset \mathbb{P}^2(1,2,3)\,:\, x^6+y^3+z^2-sxyz=0\}.
\end{align}
One can obtain a fibration using the $E_8$ fiber and a base $B$ from the following toric data: 
\begin{align}
\begin{array}{c|ccccc|cc}
\multicolumn{1}{c}{\text{div.}}&\multicolumn{5}{c}{\bar{\nu}_i^*}&l^{(e)}&l^{\bullet}\\
K_M&0&0&0&0&0&-6&0\\
D_1&&&&-2&-3&0&*\\
\vdots&&\Delta_B&&\vdots&\vdots&0&*\\
D_n&&&&-2&-3&0&*\\
E&0&0&0&-2&-3&1&-\sum*\\
2\tilde{D}_e&0&0&0&1&0&2&0\\
3\tilde{D}_e&0&0&0&0&1&3&0
\end{array}
\label{eqn:torice8}
\end{align}
In particular, fibrations of this type have a section and at most $I_1$ singularities in the fiber.

\subsection{Picard-Fuchs operators}
The periods of the holomorphic $n$-form on a Calabi-Yau $n$-fold are annihilated by a set of differential operators, the Picard-Fuchs system.
For Calabi-Yau varieties constructed as hypersurfaces in a toric ambient space it is easy to write down differential equations for which
the solution set is in general larger than that spanned by the periods.
However, in many cases the solution sets are equal and it is sufficient to study the so-called GKZ-system.
How to derive the GKZ-system from the toric data and the relation to the Picard-Fuchs system is explained e.g. in \cite{Hosono:1993qy}.

\section{Amplitudes, geometric invariants and modular forms}
\label{Amplitudes}
The topological string A-model encodes \textit{Gromov-Witten invariants}, counting holomorphic maps
\begin{align}
f:\,\Sigma_{g,\bar{p}}\rightarrow M\,,
\end{align}
from pointed curves $\Sigma_{g,\bar{p}}$ of genus $g$ into $M$.
The general formula for the virtual dimension of the moduli stack of
stable maps\footnote{A map is stable if it has at most a finite number of non-trivial automorphisms that preserve marked and nodal points.}
into a Calabi-Yau $M$ is given by
\begin{align}
\text{vir}\,\text{dim}\,\bar{M}_{g,n}(M,\beta)=(\text{dim}\,M-3)(1-g)+n\,,
\label{virtdim}
\end{align}
where $n$ is the number of marked points and we require $f_*[\Sigma]=\beta\in H_2(M)$ for $f\in\bar{M}_{g,n}(M,\beta)$ and $\Sigma$ the domain of $f$.

While for Calabi-Yau threefolds the virtual dimension is zero at all genera with $n=0$, in the case of fourfolds it is non-negative only when $g=0,1$.
A positive virtual dimension can be compensated by intersecting with classes on $M$ pulled back along
the evaluation maps
\begin{align}
\text{ev}_{i}=f(p_i):\,\bar{M}_{g,n}\rightarrow M\,,\quad i\in{0,...,n}\,.
\end{align}
On the other hand, intersecting with the pull-back of the fundamental class $[M]$ leads to vanishing invariants.
The latter property of Gromov-Witten invariants is called the Fundamental class axiom.
It follows that for fourfolds the invariants with $g\ge 2$ vanish.

We will now review the Calabi-Yau fourfold invariants for $g=0,1$ and how they are encoded in various observables of the topological A-model.

\subsection{Review of genus zero invariants}
\label{Genuszero}
From the general virtual dimension formula we find
$\text{vir}\,\text{dim}\,\bar{M}_{0,1}=2$\,,
and given $\gamma\in H^{2,2}(M,\mathbb{Z}$) we obtain well-defined invariants
\begin{align}
N_{0,\beta}(\gamma)=\int\limits_{\xi}\text{ev}_1^*(\gamma)\,,
\end{align}
with $\xi=[\bar{M}_{0,1}(M,\beta)]_{\text{virt.}}$.
From the topological string theory perspective they are encoded in the instanton part of the normalized double-logarithmic quantum periods
\begin{align}
F^{(0)}_\gamma=\text{classical}+\sum\limits_{\beta\ge0}N_{0,\beta}(\gamma)q^\beta\,.
\end{align}
In particular, the classical terms corresponding to $F^{(0)}_\gamma$ are determined by $Z^{\text{asy}}(\mathcal{O}_\gamma)$.
While the Gromov-Witten invariants are in general rational numbers, they are conjecturally related to integral \textit{instanton numbers} $n_{0,\beta}$ via
\begin{align}
\sum\limits_{\beta\ge 0} N_{0,\beta}(\gamma)q^\beta=\sum\limits_{\beta\ge 0}n_{0,\beta}(\gamma)\sum\limits_{d=1}^\infty\frac{q^{d\beta}}{d^2}\,.
\end{align}

The Gromov-Witten invariants can also be related to \textit{meeting invariants} $m_{\beta_1,\beta_2}$ \cite{Klemm:2007in}, which for $\beta_1,\beta_2\in H_2(M,\mathbb{Z})$ virtually
enumerate rational curves of class $\beta_1$ meeting rational curves of class $\beta_2$.
They are recursively defined via the following rules.
\begin{enumerate}
\item The invariants are symmetric,
\begin{align}
m_{\beta_1,\beta_2}=m_{\beta_2,\beta_1}\,.
\end{align}
\item If either $\text{deg}(\beta_1)\le0$ or $\text{deg}(\beta_2)\le0$, then $m_{\beta_1,\beta_2}=0$.
\item If $\beta_1\ne \beta_2$, then
\begin{align}
m_{\beta_1,\beta_2}=\sum\limits_{i,j}n_{0,\beta_1}(\gamma_i)\eta^{(2),ij}n_{0,\beta_2}(\gamma_j)+m_{\beta_1,\beta_2-\beta_1}+m_{\beta_1-\beta_2,\beta_2}\,,
\end{align}
where $\gamma_i\in H_V^4(M,\mathbb{Z})$ form a basis mod torsion and
\begin{align}
\eta^{(2)}_{ij}=\int\limits_M \gamma_i\cup\gamma_j\,.
\end{align}
\item If $\beta_1 = \beta_2 = \beta$, then
\begin{align}
m_{\beta,\beta} = n_{0,\beta} (c_2(T_M) ) + \sum_{i,j} n_{0,\beta}(\gamma_i) \eta^{(2),ij} n_{0,\beta}(\gamma_j) - \sum_{\beta' + \beta'' = \beta} m_{\beta',\beta''} \,.
\end{align}
\end{enumerate}
Genus one invariants for Calabi-Yau fourfolds haven been calculated for example in~\cite{Klemm:2007in,Gerhardus:2016iot}. 

\subsection{Genus one invariants}
At genus one, the virtual dimension vanishes for Calabi-Yau manifolds of any dimension.
The corresponding invariants are encoded in the holomorphic limit of the genus one free energy
\begin{align}\label{g1amp}
F^{(1)}=\text{classical}+\sum\limits_{\beta\ge 0}N_{1,\beta}q^\beta\,.
\end{align}
Assuming $h^{2,1}=0$ it has the general form
\begin{align}
F^{(1)}=\left(\frac{\chi}{24}-h^{1,1}-2\right)\log X_0+\log\det\left(\frac{1}{2\pi i}\frac{\partial z}{\partial t}\right)+\sum\limits_ib_i\log z_i-\frac{1}{24}\log\Delta\,.
\end{align}
In this expression $\chi$ is the Euler characteristic of $M$, $\Delta$ is the discriminant and $z(t)$ is the mirror map in terms of the algebraic coordinates $z$ and the flat coordinates $t$.
The coefficients $b_i$ can be fixed by the limiting behaviour of $F^{(1)}$ in the moduli space.

Assuming that the coordinates $z$ are chosen such that $z_i(t)=t_i+\mathcal{O}(t^2)$,
the large radius limit
\begin{align}\label{F1inf}
\lim\limits_{t\rightarrow\infty}F^{(1)}=-\frac{1}{24}\sum\limits_i\left(\int\limits_M c_3(M)\cup J_i\right)t_i+\text{regular}\,,
\end{align}
implies
\begin{align}\label{bconsts}
b_i=-\frac{1}{24}\int\limits_M c_3(M)\cup J_i-1\,.
\end{align}

At genus one, the conjectured relation of the Gromow-Witten numbers to integral invariants $n_{1,\beta}$ is more involved and has been worked out in \cite{Klemm:2007in}.
It involves the meeting invariants as well as the genus zero Gromov-Witten invariants and is given by
\begin{align}
\begin{split}
\sum\limits_{\beta>0}N_{1,\beta}q^\beta=&\sum\limits_{\beta>0}n_{1,\beta}\sum\limits_{d=1}^\infty\frac{\sigma(d)}{d}q^{d\beta}\\
&+\frac{1}{24}\sum\limits_{\beta>0}n_{0,\beta}\left(c_2(T_M)\right)\log(1-q^\beta)\\
&-\frac{1}{24}\sum\limits_{\beta_1,\beta_2}m_{\beta_1,\beta_2}\log(1-q^{\beta_1+\beta_2})\,.
\end{split}
\end{align}

In Appendix \ref{Gromovone} we provide genus one invariants of the one parameter fourfold geometries discussed in \cite{Bizet:2014uua}.
In the following, apart from studying the modular properties of the amplitudes, we calculate the integral invariants for
$E_8$ fibrations with bases $\mathbb{P}^3$ and $\mathbb{P}^1\times\mathbb{P}^2$. We provide some of the invariants in Appendix \ref{GromovWittensection}.
To our knowledge the latter case has not been studied in the literature before and provides further evidence supporting the conjectured
relations.

\subsection{Quasi modular forms and holomorphic anomaly equations}
\label{Quasimodular}   

In this section we further explore aspects of modularity on elliptically fibered fourfolds (with at most $I_1$ singular fibers) that has previously been observed by \cite{Haghighat:2015qdq}.
The latter authors have proven modularity of the 4-point function with all legs in the base and found modular expansions for the genus zero string amplitudes discussed in section \ref{Genuszero}.
We aim here to derive corresponding modular anomaly equations.
For the K3 case this was done in \cite{Hosono:1999qc} and for elliptic threefolds in \cite{Klemm:2012sx,Scheidegger:2012ts}.
Our strategy will be the following: We study the the generic degree $24$ hypersurface $X_{24}$ in $\mathbb{P}(1,1,1,1,8,12)$ that has been used by \cite{Haghighat:2015qdq}
to illustrate the modular structure and we find the modular anomaly equations.
We borrow the differential operators of \cite{Huang:2016SM,HKK} and find their corresponding version for CY fourfolds by comparing with the observed modular anomaly equations.
Then, we conjecture the general form of such differential quations for multiparameter families of elliptically fibered toric CY fourfolds with at most $I_1$ singularities in the fiber.
This leads to a generalized version of the modular anomaly equations that we observed for $X_{24}$.
At the end of the day we provide data for another CY fourfold supporting our conjecture. 
 
For $X_{24}$ it was found in \cite{Haghighat:2015qdq} that the instanton parts of the genus zero free energies $F^{(0)}_\gamma$ admit
an expansion
\begin{equation}\label{modexp}
F^{(0),inst}_{\gamma}(\tau,\widetilde{\underline{T}}) = \sum_{\beta \in H_{2}(B,\mathbb{Z})} F_{\gamma,\beta}^{(0),inst} (\tau) \widetilde{Q}^\beta, \quad F_{\gamma,\beta}^{(0),inst} = \Bigg( \frac{q^{\frac{1}{24}}}{\eta}\Bigg)^{12c_1(B)\cdot \beta} P_{\beta}^{(0)}(\gamma) \, ,
\end{equation}
where $P_{\beta}^{(0)}(\gamma)$ is a polynomial in the ring of quasimodular forms $\mathbb{C}[E_2,E_4,E_6]$ \cite{123mod}.
Note that an analogous ansatz can be used for the genus one string amplitudes\footnote{In this case the entries for $\gamma \in H^4_V(M,\mathbb{Z})$ are ommited.}.

For the latter case the modular weight of each polynomial coefficient $P^{(1)}_\beta$ is given by $w^{(1)}_\beta = 6 c_1(B) \cdot \beta$.
For the genus zero case we make a special distinction for the two observed kind of amplitudes. Given $\gamma \in H^4_V(M,\mathbb{Z})$ we might have
\begin{enumerate}[label=(\alph*)]
\item $F^{(0)}_{\gamma}$ transforming under ${\rm SL}(2,\mathbb{Z})$ with pure modular weight -2.
\item $F^{(0)}_{\gamma}$ transforming under ${\rm SL}(2,\mathbb{Z})$ with one component of modular weight -2 and another one of modular weight 0.
\end{enumerate}
It turns out that corresponding 4-cycles can be directly related to the basis introduced above \eqref{4cyclesbasis}.

First consider the $\pi${--vertical 4-cycles} $H^i$.
Note that we can express them as
\begin{align}
H^i = a^{ij} a^k \widetilde{D}_j \widetilde{D}_k\,,
\end{align}
which satisfies the intersection relations (\ref{4cyclesduality}), where $a^{ij}$ is the inverse
of $a_{ij}$.
Now the asymptotic part of the corresponding genus zero amplitudes follows from the computations of sections \ref{Gammaclass} and \ref{BasicPropertiesellipticfibre}, and reads
\begin{equation}\label{a-period}
F_{H^i}^{(0)} = \tau T^i+ h^i\tau +\frac{1}{2}a^{i} +F_{H^i}^{(0),inst}(q,\underline{Q} )\, .
\end{equation} 
Notice that the double logarithmic part is proportional to $\tau$ and only the $i$-th (twisted) base K\"ahler parameter $T^i$ appears in the classical part of the amplitude.
The first property is analogous to the behaviour of the periods $\partial_{\widetilde{T}^i}\mathcal{F}^{0}$ for CY threefolds, where $\mathcal{F}^0$ is the prepotential.
The second property can be satisfied by choosing a special basis of the threefold periods.
Following the same lines of the analysis that has been carried out for $\partial_{\widetilde{T}^{i}}\mathcal{F}^0$ in \cite{Huang:2016SM,HKK},
we find that the corresponding polynomials $P_\beta^{(0)}(H^i)$ for $F_{{H^{i}}}^{(0)}$ have modular weight $w^{(0)}_\beta(H^{i}) = 6 c_{1}(B)\cdot \beta -2$.
Hence we expect the full $F_{H^i}^{(0)}$ amplitudes to transform with modular weight $-2$.

On the other hand, the leading behaviour of the periods over the cycles $H_i$ is of the form
\begin{equation}\label{b-period}
F_{{H_{i}}}^{(0)} =  \frac{c_{ijk}}{2} \widetilde{T}^{j} \widetilde{T}^{k}+ \frac{1}{2}(c_{iij} -a_{ij})\widetilde{T}^{j}  + s_{i} + F_{{H_{i}}}^{(0),inst}(q,\underline{Q})\, ,
\end{equation}
where the constant $s_i$ can be determined as 
\begin{equation}
s_i = \frac{1}{24}\Big( \int c_2(B) \cdot D_i' -a_i \Big) + \frac{1}{12} (2a_i -3a_{ii} +2 c_{iii})\, .
\end{equation}
We find that $P_\beta^{(0)} (H_i) \in \widetilde{M}_{6c_{1}(B)\cdot \beta-2}(\Gamma_1 )\oplus\widetilde{M}_{6c_1(B)\cdot \beta}(\Gamma_1 )$.
Therefore $F_{H_k}^{(0)}$ belongs to the case (b).
This can be seen from the factorization of the Yukawa coupling $C_{\widetilde{T}^i \widetilde{T}^j \widetilde{T}^k \widetilde{T}^l}$ in (\ref{triplecouplings}),
which has modular weight -2. In the next section, to 
illustrate the reason for the different modular behaviour of the $F_{H^k}^{(0)}$ and $F_{{H_k}}^{(0)}$ amplitudes, we study the fourfold $X_{24}$.

As a special remark, recall the monodromy transformations $\tilde{T}:= \tilde{T}_e^{-1}$ and $\tilde{S}$ introduced in section \ref{FourierMukai}.
By introducing the factors in (\ref{modgroup}), we find that they generate the modular group.
However, $T$ and $S$ do not belong to the monodromy group of the geometry.
On the other hand, $\tilde{T}$ and $\tilde{S}$ act on the fiber parameter as the modular group, since
\begin{equation}
\tilde{T}: \tau \mapsto \tau + 1 \, , \quad \tilde{S}: \tau \mapsto -\frac{1}{\tau} \,, \quad \tilde{S}^2: \tau \mapsto \tau \, , \quad (\tilde{S} \tilde{T})^3: \tau \mapsto \tau .
\end{equation}
We note that the coordinates introduced (\ref{spcoord}) transform under $\tilde{T}$ and $\tilde{S}$ as
\begin{equation}\label{signtrans}
\tilde{T}: Q^k \mapsto (-1)^{a^k} Q^k \, ,\quad \tilde{S}: Q^k \mapsto (-1)^{a^k} Q^k \,.
\end{equation}
This has been explained already in \cite{Katz:2016SM,HKK} for CY threefolds.
We find that the same argument holds for CY fourfolds.
On the one hand, $\tilde{T}$ acts on $\tilde{T}^k$ trivially. 
On the other hand, straightforward calculations show that up to exponentially small terms $\tilde{S}$ acts as $T^k \mapsto T^k + \frac{a^k}{2}$. This leads to  (\ref{signtrans}). Moreover, note that the Dedekind eta function transforms as
\begin{equation}\label{Dedeta}
\eta^{12 a^k}(\tilde{T} \tau) = (-1)^{a^k} \eta^{12 a^k}(\tau) \, , \quad \eta^{12 a^k}( \tilde{S} \tau) = (-1)^{a^k} \tau^{6 a^k} \eta^{12 a^k} (\tau).
\end{equation} 
It follows that $Q^k$ and $\eta^{12a^k}$ are modular objects with the same multiplier system.
In particular, we can rewrite the instanton part of the string amplitudes in the coordinates \eqref{spcoord} as  
\begin{equation}\label{modnottilde}
F^{(0),inst}_\gamma = \sum_{\beta \in H_{2}(B,\mathbb{Z})} P_{\beta}^{(0)}(\gamma)\Bigg(\frac{Q^\beta}{\eta^{12c_1(B)\cdot \beta}}\Bigg)\, ,
\end{equation}
where each factor in the parenthesis transforms as a modular form of weight -$6 c_1(B) \cdot \beta$.
A similar expansion can be obtained for the genus one string amplitudes $F^{(1)}$.

\subsection{Modularity on the fourfold $X_{24}(1,1,1,1,8,12)$}

We now study the fourfold $X_{24}$ which has been introduced in \cite{Klemm:1996ts}. Its integrality and modular properties have been further discussed in \cite{Klemm:2007in,Haghighat:2015qdq}. Following the construction of section (\ref{Toricelliptic}) we pick the polytopes 
\begin{align}
\begin{split}
\Delta^{*B} & = \text{conv}(\{(-1,0,0),(0,-1,0),(0,0,-1),(1,1,1)\}),\\
\Delta^{*F} & = \text{conv}(\{(-1,0),(0,-1),(2,3)\}).
\end{split}
\end{align}
Here $\Delta^{*B}$ is the polytope for the base $\mathbb{P}^3$ and $\Delta^{*F}$ the fiber polytope for the $E_{8}$ fiber with special inner point $\nu_{3}^{*F} = (2,3)$.
We summarize the toric data in the following table which provides the points of the polytope $\Delta^*$ of $\mathbb{P}_{\Delta^{*}}$ together with the corresponding toric divisors $D_{x_{i}} = \{x_{i}=0\}$:
\begin{equation}  
 \begin{array}{rcrc|rrrrrr|rr|} 
    \multicolumn{2}{c}{\rm div.} &\multicolumn{2}{c}{\rm coord.}&\multicolumn{6}{c}{{\bar \nu}^*_i} &l^{(e)}& l^{(b)} \\ 
    K_M    && x_0 &&   1&     0&    0&    0&   0&   0&        -6&   0       \\ 
    D_1    && x   &&   1&    -1&    0&   0&    0&   0&         2&   0       \\ 
    D_2    && y   &&   1&     0&   -1&  0&     0&   0&         3&   0       \\
    E    && z   &&   1&     2&   3&   0&     0&   0&         1&  -4       \\ 
    L     && u_{1}   &&   1&     2&   3&   1&     1&   1&         0&   1      \\ 
    L    && u_{2} &&   1&     2&   3&  -1&     0&   0&         0&   1       \\
    L      && u_3 &&   1&     2&   3&   0&    -1&   0&         0&   1      \\ 
    L      && u_4 &&   1&     2&   3&   0&     0&  -1&         0&   1       \\
  \end{array} \ . 
  \label{F1case} 
\end{equation} 
We use the Sage - Mathematics Software System~\cite{sagemath} to calculate toric intersection numbers and the Mori cone.
We also provide a worksheet to illustrate the use of Sage for determining the topological invariants and the asymptotic expansions of the integral periods.
It can be downloaded from the page~\cite{URLsupp}.
The intersections of the divisors $\widetilde{D}_{b} = L$ and $\widetilde{D}_{e} = L + 4 E$ determine the constants defined in section \ref{BasicPropertiesellipticfibre},
  \begin{equation}
  a = 64, \quad a^b = 4, \quad a_b = 16, \quad a_{bb} = 4, \quad c_{bbb} = 1.
  \end{equation}
The Polytope $\Delta^*$ describes a degree 24 hypersurface $X_{24}$ given by the locus $P_\Delta$ in $\mathbb{P}_{\Delta^*} = \mathbb{P}(1,1,1,1,8,12)$. Let $X^{*}_{24}$ be the mirror manifold of $X_{24}$ defined by the locus $P_{\Delta^*}=0$  in $\mathbb{P}_{\Delta}$, where
\begin{equation}
P_{\Delta^*} = x_{0} \Big( z^6 ( \alpha_1 u_1^{24} + \alpha_1 u_2^{24} + \alpha_3 u_3^{24} + \alpha_4 u_4^{24}) + \alpha_0( u_1 u_2 u_3 u_4) xyz + \alpha_6 x^3  + \alpha_7 y^2 \Big).
\end{equation}
Here the $\alpha_i$ parametrize the complex structure of $X_{24}^{*}$.

By considering the torus action on the homogeneous coordinates of $\mathbb{P}_\Delta$, $x_i \rightarrow \lambda_a^{l_i^{(a)}} x_i$, the set of complex structure parameters can be reduced to the local coordinates for $\mathcal{M}_{cs}(W)$ given by
\begin{equation}\label{lcsv}
z^a = (-1)^{l_0^{(a)}}  \prod_{k=1}^{\vert \Delta^* \vert} \alpha_{k}^{l_k^{(a)}}, \quad a = 1, \ldots, h_{21}(W).
\end{equation}  
In particular the large complex structure limit is defined to be the point at $z=0$, this is the maximal degeneration point which corresponds to a large radius limit for the mirror manifold $M$ \cite{Batyrev:1994hm}. For the case of $X^{*}_{24}$ we have the following two large complex structure variables 
\begin{equation}
z_e = \frac{\alpha_5 \alpha_6^2 \alpha_7^3}{\alpha_0^6}, \quad z_b = \frac{\alpha_1 \alpha_2 \alpha_3 \alpha_4}{\alpha_5^4}.
\end{equation}

Essential for the B-model description is the nowhere vanishing holomorphic (4,0) form. This can be written as the residuum
\begin{equation}
\Omega(z) = \text{Res}_{P_{\Delta^*}=0} \frac{1}{P_{\Delta^{*}}(z)} \prod_i \frac{d X_i}{X_i},
\end{equation}
where $X_{i}$ are inhomogeneous coordinates on $\mathbb{P}_{\Delta}$.
Using the methods of \cite{Hosono:1993qy}, one can obtain the GKZ differential operators from the Mori cone vectors of $M$. From the GKZ operators one can extract the Picard-Fuchs operators $\mathcal{L}_i \Pi (z)=0$. Solving the latter differential equations, we obtain the periods $\Pi_\kappa(z) = \int_{\Gamma^\kappa} \Omega(z)$ in (\ref{fluxpotential}). In our present case we find the Picard-Fuchs operators
\begin{align}
\begin{split}
\mathcal{L}_{1} & = \theta_{e}(\theta_{e} - 4 \theta_{b}) -12z_{e}(6\theta_{e} -5)(6\theta_{e}-1), \\
\mathcal{L}_{2} & = \theta_b^4 - z_{b} (4\theta_b -\theta_e)(4\theta_b - \theta_e +1)(4\theta_b - \theta_e+2)(4\theta_b -\theta_e +3),
\end{split}
\end{align}
where $\theta_{a} = z^a\partial_{z^{a}}$. The components of the discriminant of these Picard-Fuchs operators are
\begin{align}
\begin{split}
\Delta_1 & = 1 -256 z_b\\
\Delta_2 & = (1-432 z_e)^4 - z_b z_e^4. 
\end{split}
\end{align}

Using (\ref{4cyclesbasis}), we can determine a basis $\{H^b, H_b\}$ of 4-cycles on $X_{24}$ given by
\begin{align}
\begin{split}
H^{b} & = \widetilde{D}_b^2 \\
   H_{b}  & = E\cdot \widetilde{D}_b.
\end{split}
\end{align}
For later convenience we introduce a special basis $\{H^b, H_b^{\circ}\}$ and refer to this as a `pure modular basis', where $H_{b}^{\circ}$ is given by
\begin{align}\label{puremod}
    H_b^{\circ} & =  H^b - 2 H_b.
  \end{align}
The respective genus zero string amplitudes in the basis $\{H^b,H_b\}$ given by (\ref{a-period}) and (\ref{b-period}) are
\begin{align}
\begin{split}
F_{H^{b}}^{(0)} & = 2\tau^2 + \tau t_{} + \tau + 2 + F_{H^{b}}^{(0),inst} (q,\widetilde{Q})\\
F_{H_b}^{(0)} & = \frac{1}{2} t_{}^2 -\frac{3}{2}t_{} + \frac{17}{12} + F_{H_b}^{(0),inst}(q,\widetilde{Q}).
\end{split}
\end{align}
Here $\tau$ and $t$ are the  K\"ahler moduli corresponding to the flat coordinates, which appear in the leading order of the mirror map of $z_{e}$ and $z_b$ respectively.
For $H_{b}^{\circ}$ the associated amplitude is given by $F_{H_b^{\circ}}^{(0)} = F_{H^b}^{(0)} - 2 F_{H_b}^{(0)}$.
In \cite{Haghighat:2015qdq} it has been observed that $F_{H^{b}}^{(0)}$ is of modular weight $k_{H^b} = -2$
while $F_{H_{b}}^{(0)}$ has a component of modular weight 0 and another of weight $-2$.
On the other hand $C_{tttt} = \eta^{\alpha\beta} \partial_{t}^2F_{\alpha}^{(0)} \partial_{t}^2 F_{\beta}^{(0)}$ has modular weight $k_{C_{tttt}}=-2$. The intersection matrix of 4-cycles $\eta'^{(2)}$ in the pure modular basis takes the form
\begin{align}
\quad
\eta^{(2)}=\begin{blockarray}{rrr}
H^b&H_b&\\
\begin{block}{(rr)r}
0&1&H^b\\
1&-4&H_b\\
\end{block}
\end{blockarray}\nonumber
\quad\longrightarrow\quad
\eta'^{(2)}=\begin{blockarray}{rrr}
H^b&H_b^{\circ}&\\
\begin{block}{(rr)r}
0&1&H^b\\
1&0&H_b^{\circ}\\
\end{block}
\end{blockarray}.\nonumber
\end{align}
Then from $C_{tttt} =2 \partial_t^2 F_{H^b}^{(0)} \partial_t^2 F_{H_b^{\circ}}^{(0)}$ it follows that $F_{H_{b}^{\circ}}^{(0)}$ is of modular weight $k_{H_b^{\circ}} = 0$.
Moreover, in \cite{Haghighat:2015qdq}, there have appeared signs of a modular anomaly equation for $F_{H^b}^{(0)}$.
We find that the periods $F_{H^b}^{(0)}$ satisfy the relation
\begin{equation}\label{a-modanomaly}
\frac{\partial F_{H^b,d}^{(0),inst}}{\partial E_{2}} = -\frac{1}{12} \sum_{s=1}^{d-1} s F_{H^b,d-s}^{(0),inst} F_{H^b,s}^{(0),inst}\,.
\end{equation}
For the case of $F_{H_b^{\circ}}^{(0)}$ we find that it does not follow the relation in (\ref{a-modanomaly}), but another kind of recursive relation given by
\begin{equation}\label{b-modanomaly}
\frac{\partial F_{H_b^{\circ},d}^{(0),inst}}{\partial E_{2}} = -\frac{1}{12d}\Bigg( \sum_{s=1}^{d-1} s^2 F_{H_b^{\circ},s}^{(0),inst} F_{H^b,d-s}^{(0),inst} + F_{H^b,d}^{(0),inst}\Bigg)\,.
\end{equation}
On the other hand, the genus one string amplitude can be easily computed from (\ref{g1amp}) and (\ref{bconsts}).
For $X_{24}$ this reads
\begin{equation}
F^{(1)} = 968 \log X^0 + \log \det \Bigg(\frac{\partial(z_e, z_b)}{\partial(\tau,t)}\Bigg) - \frac{1}{24} \log (\Delta_1 \Delta_2 )+\frac{959}{24} \log z_e +39 \log z_b \,.
\end{equation}
We find that $F^{(1)}$ follows an expansion of the form (\ref{modexp}) with polynomial coefficients $P^{(1)}_{d}$ of modular weight $w^{(1)}_d = 24d$,
i.e. $F^{(1)}$ transforms with modular weight $k_1 = 0$ as expected.
Moreover, we observe a recursive relation for $F^{(1)}$ in terms of the amplitude $F_{H^b}^{(0)}$
\begin{equation}\label{g1-modanomaly}
\frac{\partial F_d^{(1),inst}}{\partial E_2} = -\frac{1}{12} \Bigg( \sum_{s=1}^{d-1} s F_{s}^{(1),inst} F_{H^b,d-s}^{(0),inst} + \Big(\frac{5}{2}a_b + d\Big)F_{H^b,d}^{(0),inst} \Bigg)\,.
\end{equation}
As a special remark, the $\pi$-vertical period $F_{H^b}^{(0)}$ in $X_{24}$ closely resembles the quadratic logarithmic solution of the Picard-Fuchs operators in the elliptically fibered Calabi-Yau threefold given by an $E_{8}$ fibration over $\mathbb{P}^2$.
In the following section we make use of this similarity and extend it to the language of differential operators introduced in \cite{Huang:2016SM,HKK}.
We find that (\ref{a-modanomaly}), (\ref{b-modanomaly}) and (\ref{g1-modanomaly}) can be derived from such special differential relations.
Then we give a conjectural, generalized version of the modular anomaly equations for fourfolds.
In Appendix \ref{modX24} we provide data supporting the modular anomaly equations (\ref{a-modanomaly}), (\ref{b-modanomaly}) and (\ref{g1-modanomaly}).

\subsection{Derivation of modular anomaly equations}

In this section we use the approach of \cite{HKK,Huang:2016SM,Katz:2016SM} to derive modular anomaly equations for general, non-singular
elliptic Calabi-Yau fourfolds with $E_8$ fibers as described by the toric data in \eqref{eqn:torice8}.
In particular, we find recursive relations satisfied by the periods over $\pi$--vertical cycles and the genus one free energies.
On the other hand, we argue that the relation \eqref{b-modanomaly} is special to $X_{24}$ and
stems from a holomorphic anomaly equation satisfied by the 4-point couplings that we derive in \ref{sec:yukmod}. 

Recall that $z_e, z_b$ are complex structure parameters that can be expressed in terms of the mirror map as
\begin{align}
z_{e} = q\Big(1+\mathcal{O}(q,\widetilde{Q})\Big)\quad\text{and}\quad z_{b} = \widetilde{Q}\Big(1+\mathcal{O}(q,\widetilde{Q})\Big)\,.
\end{align}
When taking derivatives with respect to the Eisenstein series $E_2(q)$, we can keep either $z_b$ fixed or $t$ fixed.
In the first case one has to account for the $q$ dependence of $z_b$.
To distinguish between these operations $\mathcal{L}_{E_2(q)}$ is defined in \cite{Huang:2016SM,HKK} to be the derivative with $z_b$ held constant.
A derivative where $t$ is fixed is denoted by $\partial_{E_2(q)}$, i.e.
\begin{equation}\label{defdiff}
\mathcal{L}_{E_{2}} f: = \partial_{E_{2}(q)} f(q,z_{b}), \quad \partial_{E_{2}} f : = \partial_{E_{2}(q)} f(q,\widetilde{Q})\, .
\end{equation}
One immediately obtains the relations
\begin{equation}\label{Lsimp}
\mathcal{L}_{E_{2}}z_{b} = 0, \quad \mathcal{L}_{E_{2}} \tau = 0 \, . 
\end{equation}
In \cite{Huang:2015sta}, the following non-trivial results for the elliptic threefold $X_{18} \rightarrow \mathbb{P}^2$ have been derived 
\begin{equation}
X_{18}: \quad \mathcal{L}_{E_{2}} z_{e} = 0, \quad \mathcal{L}_{E_{2}} X^0 = 0, \quad \mathcal{L}_{E_{2}} t = \frac{1}{12}\partial_t \mathcal{F}^{(0),inst} \, .
\end{equation}

As we noted above, the asymptotic behavior of the periods over $\pi$--vertical cycles closely resembles that of the double logarithmic periods for elliptic Calabi-Yau threefolds.
Indeed we verified that for $X_{24}\rightarrow \mathbb{P}^3$ the relations
\begin{equation}\label{KKMrel}
X_{24} : \quad \mathcal{L}_{E_{2}} z_{e} = 0, \quad \mathcal{L}_{E_{2}} X^0 = 0, \quad \mathcal{L}_{E_{2}} t = \frac{1}{12} F_{H^b}^{(0),inst}\,,
\end{equation}
hold.
 Moreover, for any rational or logarithmic functions $f(z_{e},z_{b})$ and $g(X^0)$ one finds
\begin{equation}
\mathcal{L}_{E_{2}}f(z_{e},z_{b}) = 0 , \quad \mathcal{L}_{E_{2}} g(X^0) = 0 \, .
\end{equation}
We can relate the two differential operators in (\ref{defdiff}) by making use of (\ref{KKMrel}) and the chain rule to obtain
\begin{equation}\label{Lop}
\mathcal{L}_{E_{2}} f  =  \partial_{E_{2}}f+ \frac{1}{12}(\partial_{t} f) (F^{(0)}_{H^b})\, .
\end{equation}
Once again we replace $\partial_{t} \mathcal{F}^0 \leftrightarrow F_{H^b}^{(0)}$ in the analogous threefold relation and find
\begin{align}
\mathcal{L}_{E_{2}} F_{H^b}^{(0),inst} = 0\,.
\end{align}
Together these relations immediately imply the recursive relation observed in (\ref{a-modanomaly}),
\begin{equation}\label{amod2}
\partial_{E_{2}} F_{H^b}^{(0),\text{inst.}} + \frac{1}{12} F_{H^b}^{(0),\text{inst.}} \partial_{t} F_{H^b}^{(0),\text{inst.}} = 0 \, .
\end{equation}
We are now ready to generalize the discussion,

\subsubsection{Modular anomaly equations for periods over $\pi$--vertical $4$-cycles}
\label{multip}

We now consider a general non-singular elliptic Calabi-Yau fourfold $M$ with $E_8$ fiber as described by the toric data in \eqref{eqn:torice8}.
The definition of the differential operators \eqref{defdiff} can be extended to multiparameter families as
\begin{equation}\label{opmin}
\mathcal{L}_{E_2(q)} f: = \partial_{E_2(q)} f(q,\underline{z}), \quad \partial_{E_2(q)} f(q,\widetilde{\underline{Q}}) \, .
\end{equation}
Furthermore the relations \eqref{Lsimp} now read $\mathcal{L}_{E_{2}(q)} z^i =\mathcal{L}_{E_2(q)} \tau=0$. We conjecture the generalization of \eqref{KKMrel} to be given by
\begin{equation}\label{KKMmulti}
\mathcal{L}_{E_{2}} z_e =\mathcal{L}_{E_{2}} f(z_e,\underline{z}) = \mathcal{L}_{E_{2}} g(X^0 )= 0, \quad \mathcal{L}_{E_{2}} t^{i} = \frac{1}{12}  F^{(0),inst}_{H^i}\,.
\end{equation}
Note that $F^{(0),inst}_{H^i}$ on the right hand side of the last equation is singled out as the unique $\pi$--vertical period which only involves $t^i$ and $\tau$.
Using the chain rule and (\ref{KKMmulti}), $\mathcal{L}_{E_2}$ can be expressed as
\begin{equation}\label{genLf}
\mathcal{L}_{E_{2}} f = \partial_{E_{2}} f + \frac{1}{12}  (\partial_{t^{i}}f) F_{H^i}^{(0),inst}\,.
\end{equation}
Another useful relation we borrow from \cite{Huang:2016SM,HKK} by replacing a linear combination of $\partial_{\tilde{T}^i} \mathcal{F}^{(0)}$
that matches the leading asymptotic behaviour of $ F^{(0)}_{H^i}$ is
\begin{equation}\label{Lparz}
\mathcal{L}_{E_{2}} \partial_{t_{i}}z^{a}  = -\frac{1}{12} \delta^{i' j'} (\partial_{t_{i'}} z^a)( \partial_{t_{j'}} F_{H^i}^{(0),inst})\,. 
\end{equation}
We also assume $\mathcal{L}_{E_{2}} F_{H^{k}}^{(0),inst} = 0$ for the instanton contributions to (\ref{a-period}).
This determines the multiparameter version of the recursive relation \eqref{a-modanomaly} for the amplitudes $F_{H^k}^{(0)}$ associated to the $\pi$--vertical 4-cycles $H^k$
\begin{equation}
\frac{\partial F^{(0),inst}_{H^k,\beta}}{\partial E_{2}} = -\frac{1}{12} \sum_{\beta'+\beta'' = \beta}  \beta_{j}' F_{H^k,\beta'}^{(0),inst} F_{H^j,\beta''}^{(0),inst}. 
\end{equation}
We provide evidence of this relation for the geometry with base $\mathbb{P}_1\times\mathbb{P}_2$ in Appendix \ref{modX36}.
\subsubsection{Genus one modular anomaly equation}\label{g1multip}

For the same Calabi-Yau fourfold $M$ described in section \ref{multip}, we discuss now the modular anomaly equation for the genus one string amplitude.
Recall the form of $F^{(1)}$ given in (\ref{g1amp}).
Due to (\ref{opmin}), we find that $\mathcal{L}_{E_2}$ acts non-trivially only on the determinant contribution, 
\begin{equation}\label{F1mod1}
\mathcal{L}_{E_2} F^{(1)} = \mathcal{L}_{E_{2}} \log\Bigg(\text{det}\Big(\frac{\partial z^{b}}{\partial {t^a}}\Big)\Bigg) = \sum_{a,b} (\partial_{z^{b}} t^{a}) \mathcal{L}_{E_{2}}(\partial_{{t^{a}}} z^{b}) = -\frac{1}{12} \delta^{ij} \partial_{t^{i}}F_{H^j}^{(0),inst}.
\end{equation}
However, we acted on both the classical and the instanton contributions.
Denote the classical part by $P_{class}^{(1)}(\underline{t})= \sum_{a=1}^{h_{11}(M)} (b_a+1) t^a$, which is the linear polynomial appearing in (\ref{F1inf}). This gives a non-trivial contribution when acting with the differential operator $\mathcal{L}_{E_2}$ on $F^{(1)}$
\begin{align}\label{F1mod2}
\mathcal{L}_{E_{2}} F^{(1)} &= \mathcal{L}_{E_{2}} P^{(1)}_{class} + \mathcal{L}_{E_{2}}  F^{(1),inst}, 
\end{align}
where 
\begin{equation}
\mathcal{L}_{E_2} P^{(1)}_{class} = \sum_{i=1}^{h_{11}(B)} (b_{i}+1) \mathcal{L}_{E_2} t^i.
\end{equation}

Using both results (\ref{F1mod1}) and (\ref{F1mod2}) together with the expressions (\ref{genLf}) and (\ref{KKMrel}), we find the genus one modular anomaly equation for elliptically fibered Calabi-Yau fourfolds following the construction in section \ref{Toricelliptic},
\begin{equation}\label{g1modan}
\frac{\partial F^{(1),inst}_{\beta}}{\partial E_{2}} = -\frac{1}{12} \Bigg( \sum_{\beta'+\beta'' = \beta} \beta_{i}' F_{\beta'}^{(1),inst} F_{H^i,\beta''}^{(0),inst} +   \Big(\frac{5}{2}a_i + \beta_{i}\Big) F_{H^i,\beta}^{(0),inst}\Bigg).
\end{equation}
Again we provide the corresponnding data for the case that $B=\mathbb{P}_1\times\mathbb{P}_2$ in Appendix \ref{modX36} which provides a non-trivial check.
\subsubsection{4-point coupling modular anomaly equation}
\label{sec:yukmod}

From the B-model perspective the 4-point couplings $C_{pqrs}$ are rational functions in the complex structure variables $z_e,z^i$.
The A-model 4-point couplings can be expressed in the mirror coordinates $\underline{t}$ and are related to these via 
\begin{equation}\label{4coups}
C_{abcd}(\underline{t}) = \frac{1}{(X^0)^2} C_{pqrs}(\underline{z}) \frac{\partial z^p(\underline{t})}{\partial t^a} \frac{\partial z^q(\underline{t})}{\partial t^b}  \frac{\partial z^r(\underline{t})}{\partial t^c}  \frac{\partial z^s(\underline{t})}{\partial t^d}, \quad a,b,c,d = 1 ,\ldots , h_{11}(M) \, . 
\end{equation}
As we reviewed in the introduction, the 4-point coupling can be factorized in terms of the 3-point couplings $C_{ab}^\gamma$.
On the A-side the latter are derivatives of the string amplitudes  $C^\gamma_{ab} = \partial_{t^a} \partial_{t^b} F^{(0)}_\gamma$. The factorization of the 4-point function is given by 
\begin{equation}\label{fac4pt}
C_{abcd} (\underline{t}) = \partial_{t^a} \partial_{t^b} F^{(0)}_\gamma(\underline{t}) \eta^{(2),\gamma \delta} \partial_{t^c} \partial_{t^d} F^{(0)}_{\delta}(\underline{t})\, .  
\end{equation}

Now we act with $\mathcal{L}_{E_2}$ on the A-model 4-point coupling with all legs in the base, i.e. $C_{ijkl}(\tau,\underline{\tilde{T}})$, with $i,j,k,l = 1, \ldots h_{11}(B)$.
This leads to the relation
\begin{align}\label{4ptmod}
\begin{split}
\mathcal{L}_{E_2} C_{ijkl} =  -\frac{1}{12} \delta^{i'}_{j'}\Big( &C_{i'jkl} \partial_{t_{i}} F_{H^{j'}}^{(0),inst}
+ C_{ii'kl} \partial_{t_{j}} F_{H^{j'}}^{(0),inst}\\
& + C_{iji'l} \partial_{t_{k}} F_{H^{j'}}^{(0),inst}+ C_{ijki'} \partial_{t_{l}} F_{H^{j'}}^{(0),inst}\Big),
\end{split}
\end{align}
where we have used (\ref{Lparz}).
We can now insert (\ref{genLf}) to get a recursive relation of $C_{ijkl}$ with respect to the Eisenstein series $E_2$, i.e. a modular anomay equation for the 4-point coupling.

As an example we go back to the $E_{8}$ fibration over $\mathbb{P}^3$.  We apply the modular anomaly equation (\ref{4ptmod}) to $C_{tttt} \equiv C^{(4)}_b$, which reduces to 
\begin{equation}\label{YukmodX24}
\frac{\partial}{\partial{E_{2}}} C^{(4)}_b = -\frac{1}{12}[ (\partial_{t} C^{(4)}_b) F_{H^b}^{(0),inst} + 4 C^{(4)}_b( \partial_{t} F_{H^b}^{(0),inst})].
\end{equation}
It turns out that this implies the recursive relation \eqref{b-modanomaly}.
To see this we insert the factorization of $C^{(4)}_b$ given in (\ref{fac4pt}). We choose the basis $\{H^b,H_{b}^\circ\}$ introduced in (\ref{puremod}). In such a basis 
the equation we found in (\ref{YukmodX24}) can be brought into the form
\begin{equation}
\frac{\partial}{\partial E_{2}}\Big(\partial_{t}^2F_{H_{b}^\circ}^{(0),inst}\Big) \partial_{t}^2 F_{H^b}^{0,inst} = -\frac{1}{12} \Big( \partial_{t}F_{H^b}^{(0),inst} +\partial_{t}\big( F_{H^b}^{(0),inst}\cdot \partial_{t}^2 F_{H_{b}^\circ}^{(0),inst}\big) \Big) \partial_{t}^2 F_{H^b}^{(0),inst}.
\end{equation}
We can now cancel $\partial_{t}^2 F_{H^b}^{(0),inst}$ on both sides of the equation and integrate with respect to $t$. 
The result is the modular anomaly equation \eqref{b-modanomaly} satisfied by $F_{H_{b}^\circ}^{(0)}$
\begin{equation}
\frac{\partial}{\partial E_{2}} \partial_{t} F_{H_{b}^\circ}^{(0),inst} = -\frac{1}{12} \Big( F_{H^b}^{(0),inst}\partial_{t}^2 F_{H_{b}^\circ}^{(0),inst} + F_{H^b}^{(0),inst}\Big).
\end{equation}

It is immediately clear that this does not generalize to multiparameter families and periods $F_{H_i}^{(0)}$ (\ref{b-period}). We can always obtain a basis $\{H^i,H_i^{\circ}\}$ such that $F_{H_i^{\circ}}^{(0)}$ has modular weight 0. This basis only has to satisfy that $\eta^{(2)}$ is anti-block-diagonal. Then the 4-point coupling with all legs in the base is given by
\begin{equation}
C_{ijkl} = 2 \sum_{m=1}^{h_{11}(B)}\partial_{t_i} \partial_{t_j} F_{H^m}^{(0),inst} (c_{ikl} + \partial_{t_k} \partial_{t_l} F_{H_m^\circ}^{(0),inst} ), \quad i,j,k,l = 1, \ldots h_{11}(B)\,.
\end{equation}
Acting with $\mathcal{L}_{E_2}$ leads to the relation 
\begin{equation}
\mathcal{L}_{E_{2}} C_{ijkl} = \cdots + 2 \sum_{m =1}^{h_{11}(B)} (\partial_{t_{i}} \partial_{t_j} F_{H^m}^{(0),inst}) \partial_{E_2}\Big(\partial_{t_k} \partial_{t_l} F_{H_m^{\circ}}^{(0),inst}\Big), 
\end{equation}
which cannot be factorized as was possible in the case of $X_{24}$.

\section{Horizontal flux vacua for $X_{24}^*$}
\label{Vacua}  

We will now use the integral period basis for the mirror $X_{24}^*$ of $X_{24}(1,1,1,1,8,12)$ to study the admissible horizontal fluxes and the corresponding vacua.
To this end we analytically continue the basis to various special loci in the complex structure moduli space.
Note that the structure of the moduli space is similar to that of the mirror of the threefold $X_{18}(1,1,1,6,9)$ which has been studied in \cite{Candelas:1994hw}. 

Recall the defining equation of $X_{24}^*$,
\begin{equation}
z_b\,u_1^{24} + u_2^{24} + u_3^{24} + u_4^{24} + (u_1u_2u_3u_4)^6+ u_1 u_2 u_3 u_4 xy + z_e^{\frac{1}{2}}\,x^3  + y^2=0\,.
\label{eq1x24}
\end{equation}
The two components of the discriminant are given by the vanishing loci of
\begin{align}
\Delta_1=1-2^8\cdot z_b\,,\quad \Delta_2=2^{24}3^{12}\cdot z_e^4z_b-\left(1-2^43^3\cdot z_e\right)^4\,.
\end{align}

First we introduce a new set of complex structure variables by rescaling the homogeneous coordinates on $\mathbb{P}(1,1,1,1,8,12)$.
The defining equation \eqref{eq1x24} becomes
\begin{equation}
u_1^{24} + u_2^{24} + u_3^{24} + u_4^{24} + 4\,\phi\,(u_1u_2u_3u_4)^6+ 2\sqrt{3}\,\psi\, u_1 u_2 u_3 u_4 xy + x^3  + y^2=0\,,
\label{eq2x24}
\end{equation}
and the new complex structure variables $\phi,\psi$ are related to $z_e,z_b$ via
\begin{align}
z_b=\frac{1}{256}\frac{1}{\phi^4}\,,\quad z_e=\frac{1}{432}\frac{\phi}{\psi^6}\,.
\end{align}
In these variables the components of the conifold become
\begin{align}
\begin{split}
\Delta_1'=(\phi-1)(\phi+1)(1+\phi^2)\,,\quad\Delta_2'=(\phi'-1)(\phi'+1)(1+\phi'^2)\,,
\end{split}
\end{align}
where we introduced $\phi'=\phi-\psi^6$.
\begin{figure}[h!]
\centering
\includegraphics[width=.7\linewidth]{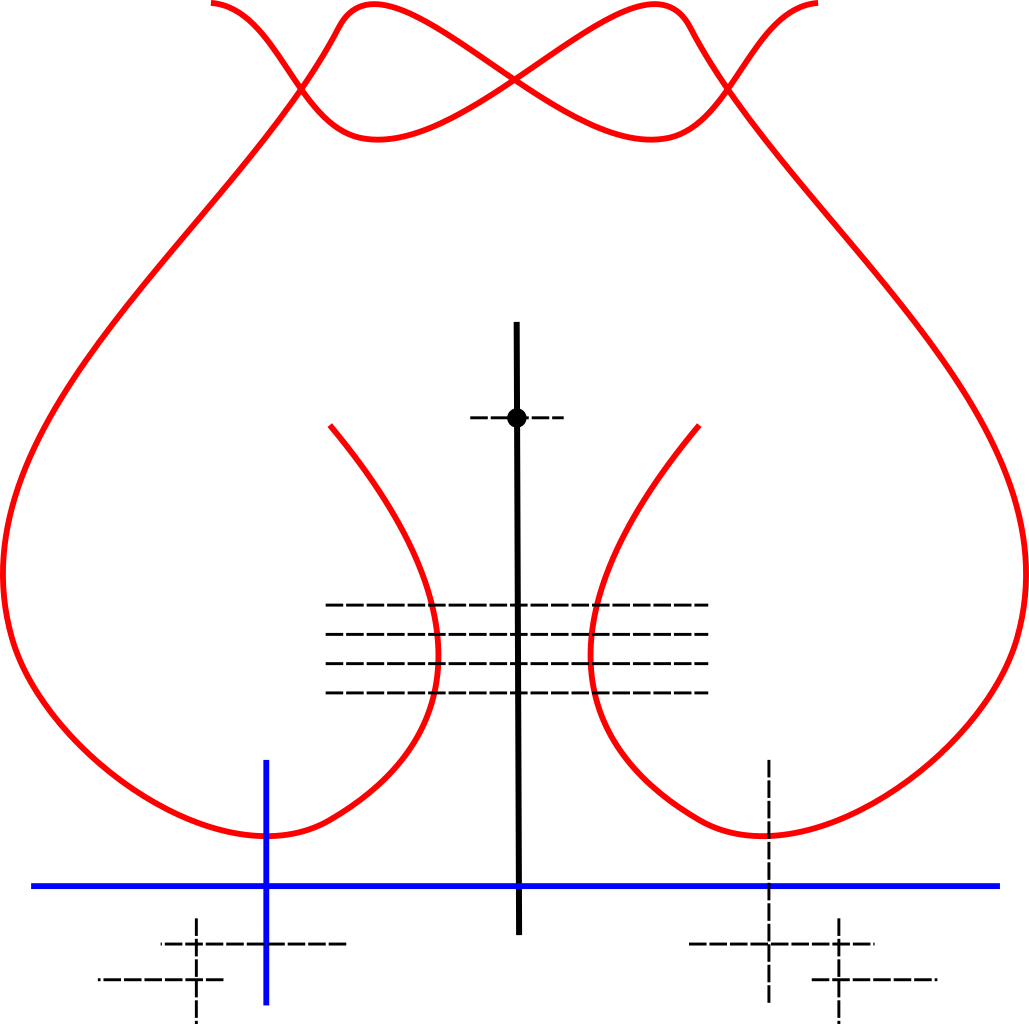}
\caption{Schematic structure of the resolved complex structure moduli space of $X_{24}^*$. The large complex structure divisors are shown in blue and the conifold components are red. Exceptional divisors resolving non-normal crossing intersections are indicated with dashed lines.}
\label{fig:resolved_moduli}
\begin{tikzpicture}[remember picture,overlay,node distance=4mm,block/.style = {draw, rectangle, minimum height=65mm, minimum width=83mm,align=center},]
\node[] at (-5.1,3.8) {$\text{LR}_2$};
\node[] at (-1.9,2.3) {$\text{LR}_1$};
\node[] at (0,9.6) {$O_1$};
\node[] at (-.3,7.8) {$P$};
\node[] at (-4,10) {$C_1$};
\node[] at (4,10) {$C_2$};
\end{tikzpicture}
\end{figure}
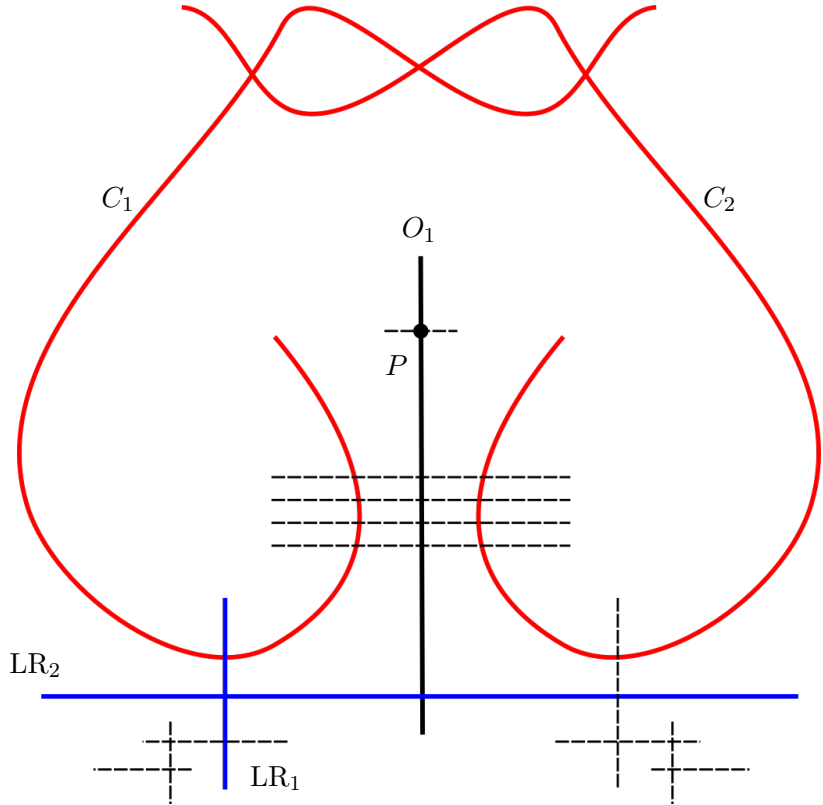

The general structure of the moduli space is sketched in figure \ref{fig:resolved_moduli}.
Note that $z_e$ and $z_b$ are the Batyrev variables and the large complex structure divisors $\text{LR}_1$ and $\text{LR}_2$ correspond
to $z_e=0$ and $z_b=0$ respectively.
On the other hand, using $\phi$ and $\psi$ as variables, both $\Delta_1'=0$ and $\Delta_2'=0$ have a forth-order tangency with $\text{LR}_2$.
Only after resolving $\text{LR}_2\cap\{\Delta_1'=0\}$ we get $\text{LR}_1$ as one of the exceptional divisors.
This is reflected in the fact that the point $\{z_e=0\}\cap\{z_b=0\}$ corresponds to a double-scaling limit in $\phi$ and $\psi$.
The two divisors that correspond to the components of the conifold are labelled with $C_1$ and $C_2$ respectively.
Furthermore, we will analyze solutions around the orbifold divisor $O_1$ that is given by $\psi=0$.

Finally note that $\Delta_1$ and $\Delta_2$ as well as $\text{LR}_1$ and $\text{LR}_2$ are exchanged under the involution 
\begin{align}
z_e=2^{-4}3^{-3}-z_e'\,,\quad z_b=\left(\frac{2^43^3z_e'}{1-2^43^3z_e'}\right)^4z_b'\,.
\end{align}
Physically this involution can be seen as the result of T-dualizing along both cycles of the fiber and
the corresponding transformation of the A-brane charges is given by $\tilde{S}$, \eqref{eqn:bridgelandX24}.
\subsection{Conifold $C_1$}
First we study the possible fluxes around $C_1\cap \text{LR}_1$. 
To this end we choose local coordinates
\begin{align}
c_1=z_b+\frac{1}{256}\,,
\end{align}
and $z_e$.
We transform and solve the Picard-Fuchs equations to obtain a vector of eight solutions with asymptotic behaviour given by
\begin{align}
\begin{split}
\Pi_{c}=&\left(
 1,\, c_1,\, z_e,\, \log \left(z_e\right),\right.
\left.\log ^2\left(z_e\right),\, \log ^3\left(z_e\right),\, \log ^4\left(z_e\right),\, c_1^{3/2}
\right)+\mathcal{O}(c^2,z^2)\,.
\end{split}
\end{align}
We demand that the leading monomial of each period is absent from the other solutions to specify the vector uniquely.

This is related to the integral basis at large complex structure via
\begin{align}
\Pi_{LR}=T_c\cdot \Pi_c\,.
\label{defTc}
\end{align}
The matrix $T_c$ can be obtained by numerical analytic continuation and is given by
{
\begin{align*}
\begin{split}
\left(
\begin{array}{cccccccc}
 f_{1,1} & f_{1,2} & f_{1,3} &f_{1,4}& \frac{54 \pi ^6 r_4^2-91}{24 \pi ^2} & r_4 & \frac{1}{6 \pi ^4} & 0 \\
f_{2,1}& \left(1+i \sqrt{2}\right) r_3 & f_{2,3} & 0 & 0 & 0 & 0 & \frac{10240 i \sqrt{2}}{3 \pi ^2} \\
 f_{3,1} & f_{3,2} & f_{3,3} &f_{3,4}& \frac{1-6 i \pi ^3 r_4}{4 \pi ^2} & -\frac{i}{3 \pi ^3} & 0 & 0 \\
 r_1+i r_2+1 & \frac{1}{4} \left(2+3 i \sqrt{2}\right) r_3 & f_{4,3} & 0 & 0 & 0 & 0 & \frac{2048 i \sqrt{2}}{\pi ^2} \\
 f_{5,1} & f_{5,2} & f_{5,3} & -\frac{3 \pi ^3 r_4+i}{2 \pi } & -\frac{1}{2 \pi ^2} & 0 & 0 & \frac{1024 i \sqrt{2}}{3 \pi ^2} \\
 \frac{2 i r_2}{3} & \frac{i r_3}{\sqrt{2}} & f_{6,3} & 0 & 0 & 0 & 0 & \frac{4096 i \sqrt{2}}{3 \pi ^2} \\
f_{7,1}& -\frac{i \left(\sqrt{2} \pi  r_3-256\right)}{8 \pi } & f_{7,3} & \frac{i}{2 \pi } & 0 & 0 & 0 & -\frac{1024 i \sqrt{2}}{3 \pi ^2} \\
 1 & 0 & 60 & 0 & 0 & 0 & 0 & 0 
\end{array}
\right)\,.
\end{split}
\end{align*}
}
Using the algebraic constraint
\begin{align}
\int\Omega\wedge\Omega=0\quad\Leftrightarrow\quad\Pi_c^T T_c^T\eta^{-1}T_c\Pi_c=0\,,
\end{align}
and the integral monodromies corresponding to $\text{LR}_1$ and $\text{C}_1$\footnote{Since we performed the analytic continuation to very high precision, the integral
monodromy matrix corresponding to $c_1\rightarrow e^{2\pi i}c_1$ is essentially determined by the numerical value.} we reduced the numerical uncertainty to five real values $r_i,\,i=1,...,5$.
Due to the size of the expressions we relegated the elements $f_{*,*}$ and the numerical values into appendix \ref{app:continuation}.

To further simplify the analysis we will move away from $z_e=0$ and introduce
\begin{align}
c_2=z_e-\frac{1}{1728}\,.
\end{align}
The corresponding vector of solutions is given by
{\renewcommand\arraystretch{1.3}
\begin{align}
\Pi_{c'}=\left(\begin{array}{c}
1-3840c_1c_2+430080c_1^2c_2\\
c_1-1920c_1^2c_2\\
c_1^2\\
c_1^3\\
c_2+32c_1c_2-29568c_2^2c_1-\frac{13216}{3}c_1^2c_2\\
c_2^2+\frac{1}{18}c_1c_2+64c_2^2c_1-\frac{40}{9}c_1^2c_2\\
c_2^3+\frac{1}{12}c_2^2c_1+\frac{1}{432}c_1^2c_2\\
c_1^{3/2}-\frac{2024}{9}c_1^{5/2}
\end{array}\right)+\mathcal{O}(c^4)\,.
\end{align}
}
This is related to the integral basis at large complex structure via
\begin{align}
\Pi_{LR}=T_c\cdot T_{c'}\cdot \Pi_{c'}\,.
\end{align}
The numerical value of $T_{c'}$ as well as those of the other continuation matrices in this section are provided in a Mathematica worksheet that can be downloaded 
from~\cite{URLsupp}.

We now obtain the monodromy action 
\begin{align}
M_c=\left(\begin{array}{cccccccc}
1&0&0&0&0&0&0&0\\
0&-9&0&20&0&-10&0&-10\\
0&0&1&0&0&0&0&0\\
0&-6&0&13&0&-6&0&-6\\
0&-1&0&2&1&-1&0&-1\\
0&-4&0&8&0&-3&0&-4\\
0&1&0&-2&0&1&1&1\\
0&0&0&0&0&0&0&1
\end{array}\right)\,,
\end{align}
on $\Pi_{LR}$ when transported along a lasso wrapping $C_1$.
Using the algebraic constraints
\begin{align}
\begin{split}
\int\Omega\wedge\Omega=0\,,\quad \int\Omega\wedge\partial_{c_1}\Omega=0\,,\quad \int\Omega\wedge\partial_{c_2}\Omega=0\,,
\end{split}
\end{align}
we find the analytic expression for $(T_cT_{c'})^T\eta^{-1}T_cT_{c'}$.
Unfortunately we are unable to solve the resulting equation for $T_{c'}$.

However, note that
\begin{align}
M_c=\mathbb{I}-\vec{v}\cdot \vec{v}^T\cdot \eta^{-1}\,,
\end{align}
where $\vec{v}=\pm\,(0,\,10,\,0,\,6,\,1,\,4,\,-1,\,0)$.
In other words, the monodromy $M_c$ corresponds to a Seidel-Thomas twist, where the charge of the shrinking brane is given by
\begin{align}
\pi_c=\vec{v}\eta^{-1}T_cT_{c'}\Pi_{c'}=\frac{2048\sqrt{2}}{3\pi^2}\left(c_1^{\frac{3}{2}}-\frac{2024}{9}c_1^{\frac{5}{2}}+\mathcal{O}(c^4)\right)\,.
\end{align}

Let us insert the topological invariants \eqref{constants} into \eqref{eqn:fmellipticmatrix} to obtain
the action of the Bridgeland type involution on $\Pi_{LR}$\,,
\begin{align}
\tilde{S}=\left(
\begin{array}{cccccccc}
 0 & -1 & 0 & 4 & 0 & -6 & 0 & -2 \\
 1 & 0 & 0 & 0 & 0 & 0 & 0 & 0 \\
 0 & 0 & 0 & -1 & 0 & 4 & 0 & -2 \\
 0 & 0 & 1 & 0 & 0 & 0 & 0 & 0 \\
 0 & 0 & 0 & 0 & 0 & -1 & 0 & 3 \\
 0 & 0 & 0 & 0 & 1 & 0 & -1 & 0 \\
 0 & 0 & 0 & 0 & 0 & 0 & 0 & -1 \\
 0 & 0 & 0 & 0 & 0 & 0 & 1 & 0 \\
\end{array}
\right)\,.
\label{eqn:bridgelandX24}
\end{align}
Then we observe
\begin{align}
\vec{v}\cdot\eta^{-1}\cdot \tilde{S}\cdot \tilde{T}_2^{-2}=(1,\,0,\,0,\,0,\,0,\,0,\,0,\,0)\,,
\end{align}
where $\tilde{T}_2$ is the monodromy corresponding to $\text{LR}_2$.
The involution exchanges $C_1$ and $C_2$ and transforms the brane vanishing at $C_1$, up to large complex structure monodromies, into a $D_8$-brane. 

To obtain a vanishing superpotential at $c_1=0$, we can turn on $n\in\mathbb{Z}$ units of flux along the cycle with period $\pi_c$.
For this to be a supersymmetric minimum we also have to check that $D_iW=0$.
In flat coordinates $t_c^i$ this condition reads
\begin{align}
(\partial_i+K_i)W=0\,,
\end{align}
where $K_i=\partial_i K$ and $K$ is the K\"ahler potential
\begin{align}
e^{-K}=\int\bar{\Omega}\wedge\Omega=\Pi_{LR}^\dagger \eta^{-1}\Pi_{LR}\,.
\end{align}

As flat coordinates we can use the normalized periods
\begin{align}
\begin{split}
t_c^1=&\frac{\Pi_{c',2}}{\Pi_{c',1}}=c_1+1920c_1^2c_2+\mathcal{O}(c^4)\,,\\
t_c^2=&\frac{\Pi_{c',5}}{\Pi_{c',1}}=c_2+32c_1c_2-\frac{13216}{3}c_1^2c_2-25728c_1c_2^2+\mathcal{O}(c^4)\,.
\end{split}
\end{align}
In terms of these, the vanishing period reads
\begin{align}
\pi_c=\frac{2048\sqrt{2}}{3\pi^2}\left[(t_c^1)^{\frac{3}{2}}-\frac{2024}{9}(t_c^1)^{\frac{5}{2}}+\mathcal{O}(t_c^4)\right]\,.
\end{align}
Using the numerical result for $T_c\cdot T_{c'}$ we find that $\partial_i K$ are regular at $c_1=0$ and therefore $D_i\pi_c\sim (t_c^1)^{i-1/2}$.

\begin{figure}[h!]
\centering
\includegraphics[width=.5\linewidth]{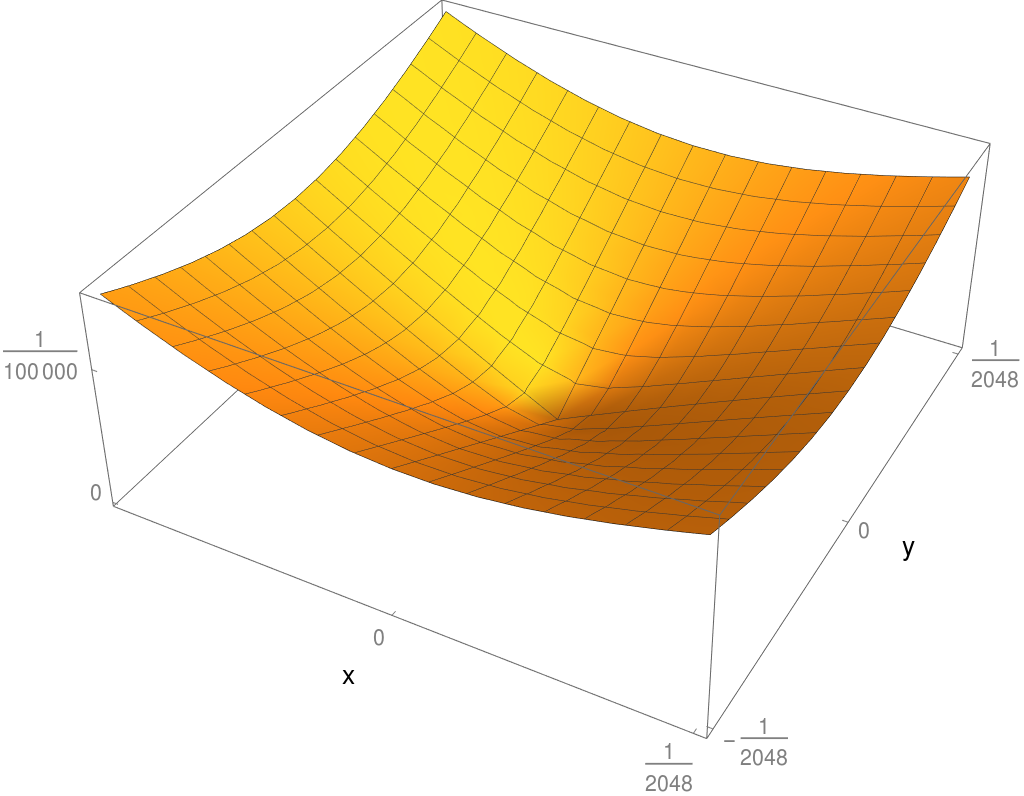}
\caption{The scalar potential generated by aligned flux, depending on the distance to the conifold $C_1$ in flat coordinates $t_c^1=x+I y,\, t_c^2=0$.}
\label{fig:scalar_potential_c1}
\end{figure}

The scalar potential is given by
\begin{align}
\begin{split}
v=&e^K\left[(D_iW)(D_{\bar{j}}\bar{W})G^{i\bar{j}}-3W\bar{W}\right]\,,
\end{split}
\end{align}
where $G^{i\bar{j}}$ is the inverse of the metric $G_{i\bar{j}}=\partial_i\partial_{\bar{j}}K$.
We restrict to $t_c^2=0$ and introduce $\text{Re}(t_c^2)=x,\,\text{Im}(t_c^2)=y$.
Then the leading terms of the scalar potential are
\begin{align}
v=&0.020174 \sqrt{x^2+y^2}+0.31715 x^2+0.31715 y^2-2.8019 x \sqrt{x^2+y^2}+\mathcal{O}(x^3,y^3)\,.
\end{align}
A plot is shown in figure \ref{fig:scalar_potential_c1}.
We checked that this is the dominant contribution at least up to order seven, where we calculated the coefficients
to a precision of twenty digits.
Deep inside the radius of convergence $|t_c^1|\approx|c_1|<1/256$ the potential is well approximated by the leading order $v\approx 0.020174 \cdot|c_1|$.
Our findings are in agreement with \cite{Bizet:2014uua} where it was argued that for Calabi-Yau fourfolds the Conifold is generically stabilized
by aligned flux.

\subsection{Orbifold $O_1$}
To expand around $O_1\cap\text{LR}_2$ we use the variables $z_b$ and
\begin{align}
o_1=\frac{1}{z_b^6}\,.
\end{align}
We find a vector of solutions to the transformed Picard-Fuchs system with leading terms
\begin{align}
\begin{split}
\Pi_{o}=&\left(o_1^{5},\,o_1^{5}\,\log\left(z_b\right),\,o_1^{5}\,\log^2\left(z_b\right),\,o_1^{5}\,\log^3\left(z_b\right),\,\right.\\
&\left.o_1,\,o_1\,\log\left(z_b\right),\,o_1\,\log^2\left(z_b\right),\,o_1\,\log^3\left(z_b\right)\right)+\mathcal{O}(o_1^{7},z)\,.
\end{split}
\end{align}
It is related to the integral basis at large complex structure via
\begin{align}
\Pi_{\text{LR}}=T_o\cdot \Pi_o\,.
\end{align}
However, in contrast to the analytic continuation matrix to the conifold, $T_o$ can be determined exactly with the help
of the Barnes integral method.
The latter has been discussed for one-parameter models in \cite{Huang:2006hq} and can be adapted to this two-parameter model.
We give the analytic expression in the Mathematica worksheet that can be found online~\cite{URLsupp}.
The monodromy acting on $\Pi_{\text{LR}}$ when transported along a lasso wrapping $O_1$ is of order six and given by
\begin{align}
M_o=\left(
\begin{array}{cccccccc}
 1 & 1 & 0 & -4 & 0 & 6 & 0 & 2 \\
 -1 & 0 & -4 & 0 & -10 & 0 & -20 & 0 \\
 0 & 0 & 1 & 1 & 0 & -4 & 0 & 2 \\
 0 & 0 & -1 & 0 & -4 & 0 & -10 & 0 \\
 0 & 0 & 0 & 0 & 1 & 1 & 0 & -3 \\
 0 & 0 & 0 & 0 & -1 & 0 & -3 & 0 \\
 0 & 0 & 0 & 0 & 0 & 0 & 1 & 1 \\
 0 & 0 & 0 & 0 & 0 & 0 & -1 & 0 \\
\end{array}
\right)\,. 
\end{align}

To analyze the possible fluxes we will again move away from the large complex structure divisor and introduce the variable
\begin{align}
o_2=z_b-\frac{1}{512}\,.
\end{align}
Solutions in the new variables are
\begin{align}
\begin{split}
\Pi_{o'}=&\left(o_1,\,o_1^{7},\,o_1^{12},\,o_1^{19},\,o_1^{5},\,o_1^{11},\,o_1^{17},\,o_1^{23}\right)+\mathcal{O}(o_1^2,o_2)\,,
\end{split}
\end{align}
We demand that the leading monomial of each period is absent from the other solutions to specify the vector uniquely.
It is related to the previous basis via
\begin{align}
\Pi_o=T_{o'}\cdot\Pi_{o'}\,,
\end{align}
where the numerical expression for $T_{o'}$ has been calculated with a precision of around fifty digits. 
\begin{figure}[h!]
\centering
\includegraphics[width=.5\linewidth]{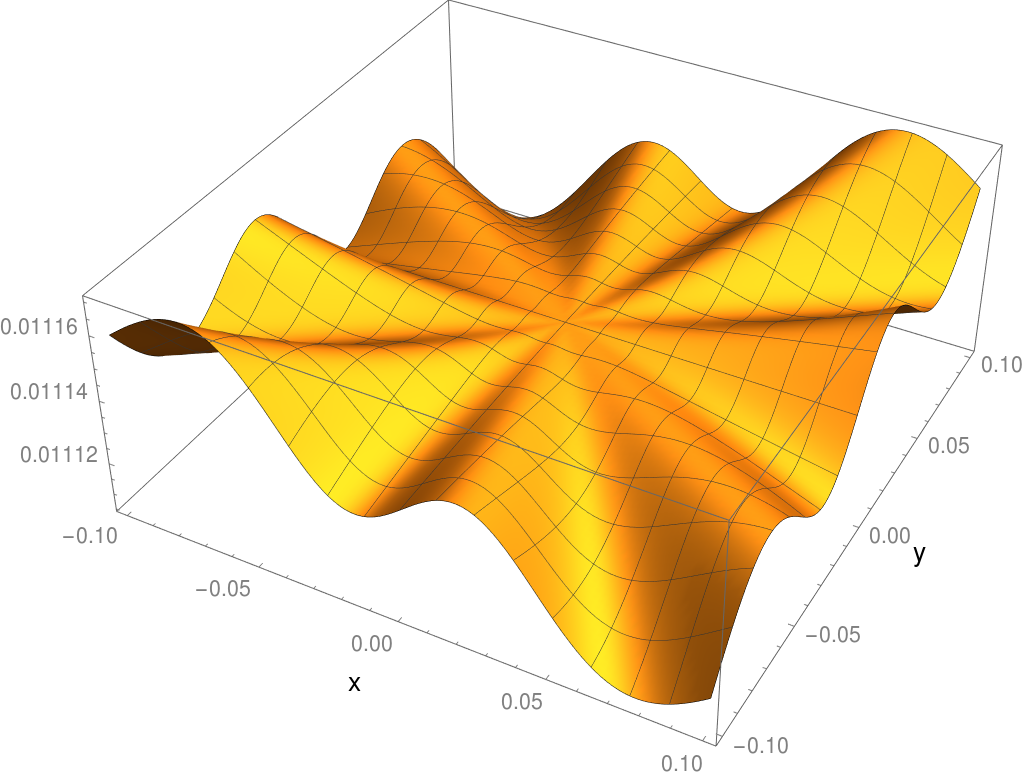}
\caption{The scalar potential generated by a generic choice of flux, depending on the distance to the orbifold $O_1$ in coordinates $o_1=x+I y,\, o_2=0$.}
\label{fig:scalar_potential_o1}
\end{figure}

From the solution vector it follows that every choice of flux leads to a vanishing superpotential at $o_1=0$.
Moreover, our numerical analysis shows that $D_{\sigma_i}W=0,\,i=1,2$ is generically satisfied at $O_1$.
If one chooses the flux superpotential
\begin{align}
\begin{split}
W=&T_{o'}^{-1}T_o^{-1}\Pi_{\text{LR},0}=-(0.237201\, -0.907908 i) o_1\\
&+(97.5605\, -9.49343 i) o_1 o_2-(24181.7+1211.32 i) o_1 o_2^2+\mathcal{O}(o^4)\,,
\end{split}
\end{align}
this leads to the scalar potential
\begin{align}
v=0.011139161558549787439+\mathcal{O}(x^2,y^2)\,.
\end{align}
in terms of $o_1=x+Iy$ at $o_2=0$.
A plot of the potential, expanded to order eleven, is shown in figure \ref{fig:scalar_potential_o1}.
Note that the radius of convergence is $o_1<216\cdot(2-2^{3/4})\approx69$.

We did a Monte Carlo scan over non-vanishing flux vectors and found that the scalar potential
was always positive at $x=y=0$.
Moreover, the behaviour close to the origin was qualitatively the same in that the gradient vanished
at $x=y=0$ but the Hessian was undefined.

We also performed an analytic continuation to the special locus $P$ where the Calabi-Yau becomes a Gepner model.
However, the behaviour of the scalar potential was qualitatively the same as for a generic point on $\mathcal{O}_1$.

\section{Conclusions and  Outlook}
\label{Conclusions}  

We described a very efficient method  to obtain the  
integral flux superpotential using the central charge formula defined in terms of the $\hat \Gamma$ 
class . This method is simple enough to  be applied to multi moduli cases. In particular if the Calabi-Yau fourfold 
is embedded in a toric ambient space  it is in general straightforward to find a basis by toric intersection calculus and the 
Frobenius method for constructing the periods at the points of maximal unipotent monodromy.  
Example calculations in Sage can be found on our homepage~\cite{URLsupp}.

We then restrict to non-singular elliptic Calabi-Yau fourfolds and study universal 
monodromies in the integral basis of the horizontal cohomology and the dual homology.  
Using this basis we provide general expressions for the monodromies corresponding to $T_i$-shifts, that act as 
$t_i\rightarrow t_i+1$ on the K\"ahler moduli.
In the derived category these correspond to the auto-equivalences 
induced by  tensoring with the line bundles  of the dual divisor. Physically this is the integral Neveu-Schwarz 
B-field shift and the action on the periods follows directly from their leading logarithms which are 
determined again by $A$-model intersection numbers.
In particular, the $T_e$-shift acts 
as the parabolic operator $T$ in ${\rm SL}(2,\mathbb{Z})$ on the fiber parameter.

More non-trivially we extend 
Bridgelands construction of an auto-equivalence of elliptic surfaces to the class of elliptic 
Calabi-Yau fourfolds with at most $I_1$ singularities in the fibers. This provides an action of the order two 
element $S$ in ${\rm SL}(2,\mathbb{Z})$ on the fiber parameter. Apart from being a non-trivial check of the integrality of  
our periods, these  auto-equivalences generate the  full ${\rm PSL}(2,\mathbb{Z})$ action  
on the elliptic parameter.  This gives rise to modular properties of the genus zero and holomorphic genus one amplitudes as well as a
holomorphic anomaly that we analyze in detail.

Let us summarize the types of  the amplitudes and the results. The virtual dimension formula  
(\ref{virtdim}) is positive for genus zero. Therefore we need a meeting condition for rational curves 
with $\gamma \in H^4(M,\mathbb{Z})$ (mod torsion) and get different amplitudes 
$F^{(0)}_\gamma(q)$ for each $\gamma$, whose geometry with respect to the fibration structure  
plays an important role.
In genus one the virtual dimension is zero and we get a universal 
amplitude $F^{(1)}$. For $g>1$ the dimension is negative and hence all higher 
genus amplitudes vanish. Finally one can also consider the modular properties of the 4-point 
functions.   The clearest situation arises for the genus zero amplitudes  associated to 
$\pi$--vertical $4$-cycles $H^{k}$ and for the genus one amplitude as well as for the 4-point functions with all legs in the base. In each case  
we get a complete and universal answer for the holomorphic anomaly equations 
which can be derived using the methods in \cite{Huang:2015sta,HKK}. 

For genus zero amplitudes over 4-cycles that are not $\pi$--vertical we observe
a modular anomaly equation only for the $E_8$ fibration over $\mathbb{P}^3$. 
However, we argue that this is a consequence of the modular anomaly equation of the 4-point function which factorizes
for two-parameter families. We also check the integrality of the 
curve counting invariants of~\cite{Klemm:2007in} at genus one for 
various new cases.

In order to study the global properties of the horizontal flux superpotential relevant for 
F-theory compactifications, we analytically continued the 
periods of the mirror $X_{24}^*$ to the following critical divisors displayed in figure \ref{fig:resolved_moduli}, 
whose symmetry  implies that we only need to consider the left half of it. We first studied the 
conifold divisor  ${C}_1$.
Here we could determine an analytic expression for the $8\times 8$ continuation matrix 
$T_c$ in (\ref{defTc}) up to five numerical coefficients\footnote{Further details about this highly non-trivial analytic continuation 
can be found at \cite{URLsupp}.}.
We also generalized  the result 
of~\cite{Bizet:2014uua} that flux along the vanishing cycle stabilizes the 
theory at this divisor. We further analyzed the possible flux superpotentials at the 
generic orbifold divisor ${O}_1$ and its special locus $P$.

The most obvious generalization of this work is to include singular elliptic fibrations. Our 
formalism for fixing the integral periods explained in section \ref{Gammaclass} will 
work essentially unchanged and with the same technical tools as long as we have 
Calabi-Yau spaces embedded into toric varieties and the resolutions of the 
singularities  can be described torically.  This will be essential to probe in a 
quantitative way the flux stabilization mechanism  of realistic F-theory vacua.    
The generalization of the construction of the Bridgeland auto-equivalence should also be possible in principle. In fact at least 
in the Calabi-Yau threefold case the results for the all genus  amplitudes 
which can be expressed in terms of Weyl-invariant Jacobi-Forms\cite{Haghighat:2014vxa,GHKKLZ} 
indicate that the affine Weyl-group of the singularity will appear as part of the 
auto-equivalences of the derived category of the $A$-model.                         
        
\section*{Acknowledgments} We would like to thank Andreas Gerhardus,  Thomas Grimm, 
Babak Haghighat, Min-Xin Huang, Hans Jockers, Amir-Kian Kashani-Poor and Sheldon Katz for discussions and 
comments. AK would like to thank Min-Xin Huang and Sheldon Katz for comments on 
the draft and the IHES for hospitality during the time when this work was initiated.
CFC would like to thank the financial support from the fellowship ``Regierungsstipendiaten CONACYT-DAAD mit Mexiko'' under the grant number 2014 (50015952)  and the BCGS for their generous support.
TS would like to thank the BCGS for their generous support as well as Amir-Kian Kashani-Poor and the ENS for hospitality
during part of the work on this project.

%%%%%%%%%%%%%%%%%%%%%%%%%%%%%%%%%%%%%%%%%%%%%%%%%%%%%%%%%%%%%%%%%%%%%%%%%%%%%%%%%%%%%%%%%%%%%%%%%%%%%%%%%%%%%%%%%%%%%%%%%%%%%%%%%%%%%%%%%%%%%%%%%%%%%%%%%%%%%%
\newpage
\appendix
\section{Appendix} 
\label{Appendix}  
\subsection{Curve counting invariants for one parameter fourfolds}
\label{Gromovone}
Here we report the genus zero and genus one curve counting invariants for nine one parameter 
fourfolds  in toric ambient spaces with generalized hypergeometric type Picard-Fuchs equations. The genus zero 
invariants agree with the ones calculated in~\cite{Bizet:2014uua}. The genus one 
invariants provide a new test for the multi covering formula derived in~\cite{Klemm:2007in}. 
Similar checks for one parameter Calabi-Yau spaces in Grassmannian  ambient spaces  with Apery type 
Picard-Fuchs operators were provided in~\cite{Gerhardus:2016iot}.     
 
\begin{center}
{\tiny
\begin{tabular}{ |c|cccc| } 
\hline
 $M  / n_{0,d}(J^2)$& $ d= 1 $  & $ 2 $ & $3 $ & $ 4 $\\
\hline
\multirow{4}{0em} {} $X_6(1^6)$&60480  &440884080 & 6255156277440 & 117715791990353760 \\ 
$X_{10}(1^5,5)$ &1582400 &791944986400 & 783617464399966400 & 031333248042176116592000 \\ 
$X_{3,4}(1^7)$ &16128  &17510976& 36449586432 & 100346754888576 \\
$X_{2,5}(1^7)$ & 24500  & 48263250 &  181688069500& 905026660335000 \\
$X_{4,4}(1^6,2)$ & 27904  & 71161472 & 354153540352 & 2336902632563200 \\
$X_{2,2,4}(1^8)$& 11776  & 7677952 & 9408504320 & 15215566524416 \\
$X_{2,3,3}(1^8)$ & 9396  & 4347594 & 3794687028& 4368985908840 \\ 
$X_{2,2,2,4}(1^9)$ & 6912  & 1919808 &988602624 &669909315456 \\
$X_{2,2,2,2,2}(1^{10})$  & 5120  & 852480 & 259476480& 103646279680 \\
\hline
\end{tabular}
}
\captionof{table}{Genus 0 invariants in  $F^{(0)}_{J^2}$   for nine  hypergeometric one 
parameter CY fourfold geometries.}
\end{center}
\begin{center}
{\tiny
\begin{tabular}{ |c|ccccc| } 
\hline
 $M  / n_{1,d}$& $ d= 1 $  & $ 2 $ & $3 $ & $ 4 $ & $5 $ \\
\hline
\multirow{4}{0em} {} $X_6(1^6)$& 0  & 0 & 2734099200 & 387176346729900 & 26873294164654597632 \\ 
$X_{10}(1^5,5)$ & 0  &30044000 & 3559247945776000 & 22569533194514770326000 & 88310003296637165555077889280 \\ 
$X_{3,4}(1^7)$ & 0  & 0 &  2813440 & 81906297984 & 1006848150400512  \\
$X_{2,5}(1^7)$ & 0  & 0 &  9058000& 845495712250 & 20201716419250520 \\
$X_{4,4}(1^6,2)$ & 0  & 1280 & 146150912 & 5670808217856 & 132534541018149888  \\
$X_{2,2,4}(1^8)$& 0  & 0 & 47104 & 4277292544 & 42843921424384\\
$X_{2,3,3}(1^8)$ & 0  & 0 & 53928& 1203128235 & 7776816583356\\ 
$X_{2,2,2,3}(1^9)$ & 0  & 0 &1024 &65526084 & 338199639552
  \\
$X_{2,2,2,2,2}(1^{10})$  & 0  & 0 & 3779200& 15090827264 & 27474707200000 \\
\hline
\end{tabular}
}
%\end{center}
\captionof{table}{Genus 1 invariants for several one parameter CY fourfold geometries.}
\end{center}
\subsection{Toric Data for $X_{36}$}    
\label{ToricData}

Here we consider the hypersurface $X_{36}$. This arises from an $E_8$ fibration over the base $B = \mathbb{P}^1 \times \mathbb{P}^2$. The base polytope $\Delta^{*B}$ of $\mathbb{P}^1\times \mathbb{P}^2$ is given by 

\begin{equation}  
 \begin{array}{rc|rrr|rr|} 
    \multicolumn{1}{c}{\rm div.}&&   \multicolumn{3}{c}{\quad{\bar \nu}^{*B}_i}  & l'^{(1)}& l'^{(2)} \\ 
    D'_0    &&              0&   0&        0&    -2  & -3   \\ 
    D'_2    &&               1&   1&         0&   0 & 0     \\ 
    D'_2     &&        -1&   0&         0&    0 & 0      \\ 
    D'_2    &&              0&   -1&         0&    0 & 1    \\
    D'_1      &&               0&   0&         1&    1 & 1     \\ 
    D'_1      &&                 0&  0&         -1&    1 & 1     \\
  \end{array} \ . 
  \label{F1case} 
\end{equation} 
Hence the polytope $\Delta^*$ corresponding to the fibration over $\Delta^{*E_8}$ is given by
\begin{equation}  
 \begin{array}{rcrc|rrrrrr|rrr|} 
    \multicolumn{2}{c}{\rm div.} &\multicolumn{2}{c}{\rm coord.}&\multicolumn{6}{c}{{\bar \nu}^*_i} &l^{(e)}& l^{(1)} & l^{(2)} \\ 
    K_M    && x_0 &&   1&     0&    0&    0&   0&   0& -6&        0    & 0   \\ 
    2\tilde{D}_e    && x   &&   1&    -1&    0&    0&   0&     0 &    2&   0     & 0  \\ 
    3\tilde{D}_e    && y   &&   1&     0&   -1&     0&   0&     0&     3&   0     & 0 \\
    E    && z   &&   1&     2&   3&   0&        0&        0&  1&  -2     & -3  \\ 
    \tilde{D}_2     && u_{1}   &&   1&     2&   3&        1&   1&   0&      0&   0 &1      \\ 
    \tilde{D}_2    && u_{2} &&   1&     2&   3&      -1&   0&      0&   0&   0   & 1    \\
    \tilde{D}_2      && u_3 &&   1&     2&   3&       0&   -1&      0&   0&   0   & 1   \\ 
    \tilde{D}_1      && u_4 &&   1&     2&   3&        0&  0 &      1&   0&   1    & 0    \\
    \tilde{D}_1      && u_5 &&   1&     2&   3&        0&  0&   -1&      0&   1     &0   \\
  \end{array} \ . 
  \label{F1case} 
\end{equation} 
The intersections among divisors lead to the constants in  (\ref{constants}),
\begin{align}
\begin{split}
\quad c_{ijk} & = \begin{cases}  c_{122} = c_{212} = c_{221} =  1 \, ,  \\ 0 \text{ otherwise ,} \end{cases} \\
a = 54, \quad 
a^i  = &\left(\begin{array}{c}
2\\
3\\
\end{array}\right)\,,
\quad 
a_i  =\left(\begin{array}{cc}
9 & 12\\
\end{array}\right)\,,
\quad 
a_{ij} = \left(\begin{array}{cc}
0 & 3 \\
3 & 2\\
\end{array}\right)\,.
\end{split}
\end{align}
The Picard-Fuchs equations read 
\begin{align}
\begin{split}
\mathcal{L}_1 & = \theta_e (\theta_e -2\theta_1 -3 \theta_2 )-12 z_e (6\theta_e + 5 )(6\theta_e + 1) \, , \\
\mathcal{L}_2 & = \theta_1^2 - z_1 (\theta_e -2 \theta_1 -3 \theta_2)(\theta_e -2\theta_1 -3\theta_2 -1)\, ,\\
\mathcal{L}_3 & = \theta_2^3 - z_2(\theta_e -2\theta_1 -3\theta_2)(\theta_e-2\theta_1 -3\theta_2 -1)(\theta_e-2\theta_1-3\theta_2-2 )\, .
\end{split}
\end{align}
Here the coordinates $z^a$ are determined by (\ref{lcsv}). The discriminants of the Picard-Fuchs equations are given by
\begin{align}
\begin{split}
\Delta_1 &= (-1 + 4 z_1)^3 - 54 (1 + 12 z_1) z_2 - 729 z_2^2\,, \\
\Delta_2 &= -\Big[-1 + 864 z_e \Big(1 + 216 z_e (-1 + 4 z_1)\Big)\Big]^3 +
 4738381338321616896 z_e^6 z_2^2 \nonumber \\
& \quad + 4353564672 z_e^3 (-1 + 432 z_e) \Big[1 +
    864 z_e \Big(-1 + 216 z_e (1 + 12 z_1)\Big)\Big] z_2 \,.
\end{split}
\end{align}
Using the choice of basis for 4-cycles in (\ref{4cyclesbasis}) we obtain
\begin{equation}\label{ellbasis}
H_{1} = E \cdot \tilde{D}_1 , \quad H_{2} = E \cdot \tilde{D}_2   , \quad H^{1} = \tilde{D}_2^2 , \quad H^{2} =\tilde{D}_1 \tilde{D}_2 \, . 
\end{equation}
Hence the (\ref{4cyclesduality}) intersections follow as
\begin{align}
\eta^{(2)} = \left(\begin{array}{cccc}
0 & -3 & 1 & 0\\
-3 & -2 & 0 & 1\\
1 & 0 & 0 & 0\\
0 & 1 & 0 & 0
\end{array}\right)\,.
\end{align}
The genus zero amplitudes in the basis (\ref{ellbasis}) read
\begin{align}
\begin{split}
F_{H_1}^{(0)} & = \frac{1}{2}t_2^2 -\frac{3}{2} t_2 +\frac{5}{4} + F_{H_1}^{(0),inst}(q_e,\widetilde{Q}_1,\widetilde{Q}_2) \, , \\
F_{H_2}^{(0)} & = t_1 t_2 - t_1 - t_2 +\frac{5}{4} +  F_{H_2}^{(0),inst}(q_e,\widetilde{Q}_1,\widetilde{Q}_2) \, , \\
F_{H^1}^{(0)} & = \tau^2 + \tau t_1+1  + F_{H^1}^{(0),inst}(q_e,\widetilde{Q}_1,\widetilde{Q}_2) \, , \\
F_{H^2}^{(0)} & = \frac{3}{2} \tau^2 + \tau t_2 +\frac{1}{2} \tau +\frac{3}{2} + F_{H^2}^{(0),inst}(q_e,\widetilde{Q}_1,\widetilde{Q}_2) \, .
\end{split}
\end{align}
In Appendix \ref{modX36} we show some of the instanton expansions of the above expressions in terms of quasi-modular forms. Note that we make use of a `pure modular' basis - as in the case of $X_{24}$ - to compute the modular weight zero components of the $F_{H_i}^{(0)}$ periods. We define such a basis as follows 
\begin{align}
\begin{split}
F_{H^\circ_1}^{(0)} & \equiv F_{H_1}^{(0)} + \frac{3}{2} F_{H^2}^{(0)} \, , \\ 
F_{H^\circ_2}^{(0)} & \equiv F_{H_2}^{(0)} + F_{H^2}^{(0)} + \frac{3}{2} F_{H^1}^{(0)} \, .
\end{split}
\end{align}
The second Chern class of $X_{36}$ can be written in terms of the basis (\ref{ellbasis}) as 
\begin{equation}\label{c2cyc}
c_2 (T_{X_{36}}) = 24 H_1 + 36 H_2 + 102 H^1 + 138 H^2 \,. 
\end{equation}
We compute the genus zero Gromov-Witten invariants of (\ref{c2cyc}) in Appendix \ref{GromovWittensection}.
Further constants related to the Chern classes are
\begin{align}
\begin{split}
\quad b_1& = -\frac{1}{24}c_3(T_{X_{36}}) \cdot \tilde{D}_1 -1 = \frac{43}{2}, \quad b_2 = -\frac{1}{24}c_3(T_{X_{36}})\cdot \tilde{D}_2 -1 = 29\, , \\
 b_e & = -\frac{1}{24}c_3(T_{X_{36}})\cdot \tilde{D}_e -1 = \frac{539}{4}, \quad  \chi = 19728  \, .
\end{split}
\end{align}
This leads to the genus one amplitude 
\begin{equation}
F^{(1)} =  \frac{543}{4} \tau + \frac{45}{2} t_1 + 30 t_2 + F^{(1),inst}(q_e,\widetilde{Q}_1, \widetilde{Q}_2)\,,
\end{equation}
where we give part of the expansion of $F^{(1),inst}$ in terms of quasi-modular forms in Appendix \ref{modX36}.

\subsection{Geometric Invariants for $X_{36}$} \label{GromovWittensection}

\begin{center}
{\small
\begin{tabular}{ |c|ccccc| } 
\hline
 $n_{0,(0,d_1,d_2)}(H_1)$& $d_{2}=0$  & $1$ & $2$ & $3$ & $4$ \\
\hline
\multirow{4}{0em} {} $d_1=0$ &$*$&-9&36&-243&2304 \\

1&9&-153&2745&-49734&904500 \\

2&0&-738&43506&-1719756&56117574 \\

3&0&-2250&353916&-27555633&1515365226 \\

4&0&-5355&1951704&-277450434&24502800744 \\
  \hline
 $n_{0,(1,d_1,d_2)}(H_1)$& $d_{2}=0$  & $1$ & $2$ & $3$ & $4$ \\
\hline
\multirow{4}{0em} {} $d_1=
0$&540&2160&-13500&138240&-1698840 \\

1&-1620&55080&-1456380&34833240&-786936060 \\

2&0&320760&-27424980&1396005840&-55422152100 \\

3&0&1090800&-252097380&25003580040&-1654348658580 \\

4&0&2786400&-1521167040&274895998560&-29038118214600 \\
\hline
 $n^{}_{0,(2,d_1,d_2)}(H_1)$& $d_{2}=0$  & $1$ & $2$ & $3$ & $4$ \\
\hline
\multirow{4}{0em} {} $d_{1}=0$&1080&-143370&2298240&-35363790&578799000 \\

1&249480&-11734470&409114800&-12410449830&342447273720 \\

2&-3240&-74598570&9085010220&-583905569940&27847911802680 \\

3&0&-271666710&92772238680&-11648976938100&920958991711200 \\

4&0&-731942730&605426932980&-139049122837500&17515925402297760 \\
\hline
\end{tabular}
}
\end{center}
\captionof{table}{Genus 0 invariants associated to $H_1$ of $X_{36}$ for degree $d_e =0,1,2$ of the elliptic parameter.}

\begin{center}
{\small
\begin{tabular}{ |c|ccccc| } 
\hline
 $n_{0,(0,d_1,d_2)}(H_2)$& $d_{2}=0$  & $1$ & $2$ & $3$ & $4$ \\
\hline
\multirow{4}{0em} {} $d_1=0$&$*$&-24&114&-864&8808 \\
1&6&-192&4440&-93744&1898622 \\
2&0&-744&55050&-2528040&92087760 \\
3&0&-2040&390744&-34977312&2139264666 \\
4&0&-4560&1973472&-318919680&31152820512 \\
 \hline
 $n_{0,(1,d_1,d_2)}(H_2)$& $d_{2}=0$  & $1$ & $2$ & $3$ & $4$ \\
\hline
\multirow{4}{0em} {} $d_1=
0$ &720&6120&-43920&495360&-6528960 \\
1&-720&67680&-2349360&65718720&-1654942320 \\
2&0&314280&-34350480&2043688320&-90803818800 \\
3&0&961920&-274751280&31523616000&-2326758388560 \\
4&0&2313000&-1517061600&313418304000&-36732061356480 \\
\hline
 $n^{}_{0,(2,d_1,d_2)}(H_2)$& $d_{2}=0$  & $1$ & $2$ & $3$ & $4$ \\
\hline
\multirow{4}{0em} {} $d_{1}=0$ &1440&-1036800&8217540&-131045040&2264001480 \\
1&0&-13718160&660289320&-23552058960&724510733760 \\
2&-1440&-69796080&11223041760&-851198459760&45595230845400 \\
3&0&-230700960&99434663640&-14568373007280&1290110994869760 \\
4&0&-588578400&593689222980&-157013407044000&22030115559925320 \\
\hline
\end{tabular}
}
\end{center}
\captionof{table}{Genus 0 invariants associated to $H_2$ of $X_{36}$ for degree $d_e =0,1,2$ of the elliptic parameter.}

\begin{center}
{\small
\begin{tabular}{ |c|ccccc| } 
\hline
 $n_{0,(0,d_1,d_2)}(H^1)$& $d_{2}=0$  & $1$ & $2$ & $3$ & $4$ \\
\hline
\multirow{4}{0em} {} $d_1=0$ &$*$&6&-30&234&-2424 \\
1&0&30&-870&20196&-431874 \\
2&0&84&-8682&460512&-18225348 \\
3&0&180&-51600&5535630&-376340394 \\
4&0&330&-224112&44650908&-4939206672 \\
 \hline
 $n_{0,(1,d_1,d_2)}(H^1)$& $d_{2}=0$  & $1$ & $2$ & $3$ & $4$ \\
\hline
\multirow{4}{0em} {} $d_1=
0$ &0&-1800&12240&-138240&1833120 \\
1&0&-12240&493920&-14789520&388121760 \\
2&0&-39960&5793120&-389357280&18571150800 \\
3&0&-93600&38578320&-5210146800&422999503920 \\
4&0&-181800&182164320&-45722836800&6013372484160 \\
\hline
 $n^{}_{0,(2,d_1,d_2)}(H^1)$& $d_{2}=0$  & $1$ & $2$ & $3$ & $4$ \\
\hline
\multirow{4}{0em} {} $d_{1}=0$ &0&377460&-2483820&38068380&-651769560 \\
1&0&2668140&-147669480&5533834140&-175351411440 \\
2&0&9566100&-1999807560&168934134600&-9623706319080 \\
3&0&24142860&-14689968840&2502807844680&-241840328961600 \\
4&0&49469940&-74741749380&23760824553000&-3714780571613640 \\
\hline
\end{tabular}
}
\end{center}
\captionof{table}{Genus 0 invariants associated to $H^1$ of $X_{36}$ for degree $d_e =0,1,2$ of the elliptic parameter.}

\begin{center}
{\small
\begin{tabular}{ |c|ccccc| } 
\hline
 $n_{0,(0,d_1,d_2)}(H^2)$& $d_{2}=0$  & $1$ & $2$ & $3$ & $4$ \\
\hline
\multirow{4}{0em} {} $d_1=0$ &0&3&-12&81&-768 \\
1&-3&51&-915&16578&-301500 \\
2&0&246&-14502&573252&-18705858 \\
3&0&750&-117972&9185211&-505121742 \\
4&0&1785&-650568&92483478&-8167600248 \\
 \hline
 $n_{0,(1,d_1,d_2)}(H^2)$& $d_{2}=0$  & $1$ & $2$ & $3$ & $4$ \\
\hline
\multirow{4}{0em} {} $d_1=
0)$&0&-1080&5400&-51840&617760 \\
1&720&-21240&537120&-12547080&279335520 \\
2&0&-116640&9797760&-492862320&19402918200 \\
3&0&-386640&88556760&-8722465560&574019167320 \\
4&0&-973800&528827040&-95139102240&10012773524400 \\
\hline
 $n^{}_{0,(2,d_1,d_2)}(H^2)$& $d_{2}=0$  & $1$ & $2$ & $3$ & $4$ \\
\hline
\multirow{4}{0em} {} $d_{1}=0$ &0&143370&-1149120&15155910&-231519600 \\
1&424332&4966110&-164102760&4798354950&-129069932760 \\
2&1440&29183490&-3444863940&217072351980&-10203218591040 \\
3&0&101974950&-34165780560&4236009888780&-331677657148320 \\
4&0&267877530&-218885967900&49833948532500&-6234396678989640 \\
\hline
\end{tabular}
}
\end{center}
\captionof{table}{Genus 0 invariants associated to $H^2$ of $X_{36}$ for degree $d_e =0,1,2$ of the elliptic parameter.}

\begin{center}
{\small
\begin{tabular}{ |c|ccccc| } 
\hline
 $n_{1,(0,d_1,d_2)}$& $d_{2}=0$  & $1$ & $2$ & $3$ & $4$ \\
\hline
\multirow{4}{0em} {} $d_{1}=0$& $*$ &$0$ & $0$ & $70$ & -1602 \\ 
$1$ &$0$ & $0$ & $0$ & -990 &  52884 \\ 
$2$ & 0 & 0 & 1161 & -183402 & 12496941 
  \\
$3$ &0& 0& 15174 & -4442538 & 
487139904 \\
$4$& 0& 0&110151 & 
  -56477430 & 199225723852 \\
  \hline
 $n_{1,(1,d_1,d_2)}$& $d_{2}=0$  & $1$ & $2$ & $3$ & $4$ \\
\hline
\multirow{4}{0em} {} $d_{1}=0$& -18 & 36 & -90 & -30744 & 706572 \\ 
$1$ & -18 & 288 & -5166 & 698400 &  -43754310 \\ 
$2$ & 0 & 972 & -681210& 138997944 & -11571378390
  \\
$3$ &0& 2304 & -10098990 & 3786456528 & 
-504941463486 \\
$4$& 0& 4500 & -80360496 & 
  52885199952 &  -218756626565280 \\
\hline
 $n_{1,(2, d_1,d_2)}$& $d_{2}=0$  & $1$ & $2$ & $3$ & $4$ \\
\hline
\multirow{4}{0em} {} $d_{1}=0$& 18 & -11772 & 47052 & 6547608 & -221005710 \\ 
$1$ & 4266 & -123228 & 2995704 & -257783256 &  18293975928 \\ 
$2$ &18 & -498420 &  208221228 & -54043640640 & 5483304374166 
  \\
  
$3$ &0&  -1313172 & 3364230240 &  -1643238648792 & 
265936355246088\\
$4$& 0& -2743308 & 965359376676 & 
  -273025875142044 & 36253634952195918 \\
\hline
\end{tabular}
}
\end{center}
\captionof{table}{Genus 1 invariants of $X_{36}$ for degree $d_e =0,1,2$ of the elliptic parameter.}

{\footnotesize 
\begin{center}
\begin{tabular}{ |c|ccccccccc| } 
\hline
$m_{\beta_1,\beta_2}$&$\beta_2$&&&&&&&&\\
$\beta_1$& $(0,0,1) $  & $(0,1,0)$ & $(1,0,0)$ & $(0,0,2)$ & $(0,1,1)$ & $(0,2,0)$ & $(1,0,1)$ & $(1,1,0)$ & $(2,0,0)$ \\
\hline
\multirow{4}{0em} {} $(0,0,1)$& -180 & 378 & 72 & -1422 & 0 & 5400 & 36720 & -11880 & 10800 \\ 
$(0,1,0)$ &  & -2016 & -342 & 6894 &  0 & -24840 & -138240 & 57240 & -49680 \\ 
$(1,0,0)$ &  &  & 0 & 648 & 0 & -2160 & -18360 & 0 & -4320 
  \\
$(0,0,2)$ & &  &  & -15012 & 0 & 52920 & 376920 & -85320 & 105840  \\
$(0,1,1)$&  &  &  & 
   & 0 & 0 & 0 & 0 & 0\\
$(0,2,0)$ &  & &  &  &   & 38800 & -1744200 & 516240 & 38880  \\ 
$(1,0,1)$ &  & &  &  &  &  & -7069680 & 3656800 & -3499200
  \\
$(1,1,0)$  &  & &  &  &  &  &  & 38800 & 1036800 \\
$(2,0,0)$ &  & &  &  &  &  &  &  & 38800\\
\hline
\end{tabular}
\end{center}}
\captionof{table}{Meeting invariants for $X_{36}$.}

\subsection{Modular expressions for $X_{24}(1,1,1,1,8,12)$ }\label{modX24}

{\small
\begin{align}
F^{(0),inst}_{H^b} & = P_{22}^{(0)}(H^b)\Bigg( \frac{q_e^2}{\eta^{48}} \widetilde{Q}\Bigg) +  P_{46}^{(0)}(H^b)\Bigg( \frac{q_e^2}{\eta^{48}} \widetilde{Q}\Bigg)^2 + P_{70}^{(0)}(H^b)\Bigg( \frac{q_e^2}{\eta^{48}} \widetilde{Q}\Bigg)^3 + \cdots \\
P_{22}^{(0)}(H^b) & = - \frac{5}{18} E_4 E_6 (35 E_4^3 + 37 E_6^2) \nonumber \\
P_{46}^{(0)}(H^b) & = -\frac{1}{12}E_2 \Big(P^{(0)}_{22}(H^b) \Big)^2  -\frac{5 E_4 E_6}{2985984}
\Big(29908007 E_4^9 + 207234483 E_4^6 E_6^2 
    \nonumber \\
     & \quad  + 208392741 E_4^3 E_6^4 + 27245569 E_6^6\Big) \nonumber \\
P_{70}^{(0)}(H^b) & = \frac{1}{1671768834048}5 E_4 E_6\Bigg(-129361397672887 E_4^{15} -
   2336995567194997 E_4^{12} E_6^2 \nonumber \\
   & \quad - 8349302045771014 E_4^9 E_6^4 -
   8287506676944650 E_4^6 E_6^6 - 2198284344978035 E_4^3 E_6^8 \nonumber \\
  & \quad -104063870681681 E_6^{10} -74649600 E_2^2 E_4^2 E_6^2 \Big(35 E_4^3 + 37 E_6^2\Big)^3 \nonumber \\
 &\quad - 38880 E_2 E_4 E_6 \Big(35 E_4^3 + 37 E_6^2\Big) \Big(29908007 E_4^9 +
      207234483 E_4^6 E_6^2 \nonumber \\
      & \quad+ 208392741 E_4^3 E_6^4 + 27245569 E_6^6\Big)\Bigg) \nonumber\\
      %%%%%%%%%%%%%%%%%%%%%%%
F^{(0),inst}_{H^\circ_b} & = P_{0}^{(0)} (H^\circ_b)+ P_{24}^{(0)}(H^\circ_b)\Bigg( \frac{q_e^2}{\eta^{48}} \widetilde{Q}\Bigg) +  P_{48}^{(0)}(H^\circ_b)\Bigg( \frac{q_e^2}{\eta^{48}}\widetilde{Q}\Bigg)^2 + P_{72}^{(0)}(H^\circ_b)\Bigg( \frac{q_e^2}{\eta^{48}} \widetilde{Q}\Bigg)^3 + \cdots \\
P_{0}^{(0)}(H^\circ_b) & = 960 \sum_{d=1}^{\infty} \frac{\sigma_3(d)}{d^2} q_e^d \, , \nonumber \\ 
P_{24}^{(0)}(H^\circ_b)& = \frac{5}{10368} \Big(10321 E_4^6 + 1680 E_2 E_4^4 E_6 + 59182 E_4^3 E_6^2  + 1776 E_2 E_4 E_6^3 + 9985 E_6^4\Big) \nonumber \\
P_{48}^{(0)}(H^\circ_b) & = \frac{5}{13759414272}\Bigg(34974695189 E_4^{12} +
   955855257580 E_4^9 E_6^2 + 2375228903358 E_4^6 E_6^4 \nonumber \\
  &\quad  +  958823179372 E_4^3 E_6^6 + 33221332181 E_6^8 +
   737280 E_2^2 E_4^2 E_6^2 \Big(35 E_4^3 + 37 E_6^2\Big)^2 \nonumber \\ 
  & \quad +  576 E_2 E_4 E_6 \Big(19602269 E_4^9 + 134498081 E_4^6 E_6^2 +
      137176487 E_4^3 E_6^4 + 18933723 E_6^6\Big)\Bigg) 
 \nonumber \\
 P_{72}^{(0)}(H_b^\circ) & = \frac{5}{12999674453557248} \Bigg(169868512046891311 E_4^{18} +
   10991441298020921814 E_4^{15} E_6^2 \nonumber \\
    & \quad +81579781878072712593 E_4^{12} E_6^4 +
   152135959047477825460 E_4^9 E_6^6 \nonumber \\
   & \quad+ 81740383608791276385 E_4^6 E_6^8 +
   10747159517985301398 E_4^3 E_6^{10}\nonumber \\
   & \quad + 154580302495588543 E_6^{12} +
   16124313600 E_2^3 E_4^3 E_6^3\Big(35 E_4^3 + 37 E_6^2\Big)^3  \nonumber \\
   & \quad +   8398080 E_2^2 E_4^2 E_6^2 \Big(35 E_4^3 + 37 E_6^2\Big) \Big(44357407 E_4^9 +
      305364363 E_4^6 E_6^2 + 309961101 E_4^3 E_6^4 \nonumber 
        \end{align}
      \newpage
      \begin{align}
    & \quad   + 42023369 E_6^6\Big) +  432 E_2 E_4 E_6 \Big(188980289153801 E_4^{15} +
      4041754304722571 E_4^{12} E_6^2 \nonumber \\ 
      & \quad + 14159768366734202 E_4^9 E_6^4 +
      14323384784691190 E_4^6 E_6^6 + 4070877046013005 E_4^3 E_6^8 \nonumber \\
      & \quad +
      171728558335663 E_6^{10}\Big)\Bigg) \nonumber
      \end{align}
%%%%%%%%%%%%%
\begin{align}
F^{(1),inst} & = P_{0}^{(1)} +  P_{24}^{(1)}\Bigg( \frac{q_e^2}{\eta^{48}} \widetilde{Q}\Bigg) +  P_{48}^{(1)}\Bigg( \frac{q_e^2}{\eta^{48}}\widetilde{Q}\Bigg)^2 + P_{72}^{(1)}\Bigg( \frac{q_e^2}{\eta^{48}} \widetilde{Q}\Bigg)^3 + \cdots  \\
P_{0}^{(1)} & = -2 \Big(\frac{\chi}{24} -h_{11}\Big) \sum_{d =1} \frac{\sigma_1(d)}{d} q_e^d \,,\nonumber \\
P_{24}^{(1)} & = \frac{5}{5184} \Big(-10321 E_4^6 + 34440 E_2 E_4^4 E_6 -
   59182 E_4^3 E_6^2 + 36408 E_2 E_4 E_6^3 - 9985 E_6^4 \Big) \nonumber \\
   P_{48}^{(1)} & =  \frac{5}{1719926784}\Bigg(-8718461011 E_4^{12} -
   238460285300 E_4^9 E_6^2 - 592848334770 E_4^6 E_6^4 \nonumber \\ 
&   \quad -239525096180 E_4^3 E_6^6 - 8301513619 E_6^8 +
   7649280 E_2^2 E_4^2 E_6^2 (35 E_4^3 + 37 E_6^2)^2 \nonumber \\
& \quad   +   96 E_2 E_4 E_6 \Big(599169347 E_4^9 + 4155664383 E_4^6 E_6^2 +       4173110841 E_4^3 E_6^4 + 542601349 E_6^6\Big)\Bigg) \nonumber \\
P_{72}^{(1)} & = \frac{5}{2166612408926208} \Bigg(-54494943725199823 E_4^{18} -
   3526301098569327294 E_4^{15} E_6^2 \nonumber \\
   & \quad -26187494142167356137 E_4^{12} E_6^4 -
   48905698228868539588 E_4^9 E_6^6 - \nonumber \\
 &\quad  26341691595249846705 E_4^6 E_6^8
 - 3475678553808910878 E_4^3 E_6^{10} \nonumber \\ &\quad - 50493219640852471 E_6^{12} +
   337714790400 E_2^3 E_4^3 E_6^3 \Big(35 E_4^3 + 37 E_6^2\Big)^3  \nonumber \\
   & \quad +
   1399680 E_2^2 E_4^2 E_6^2 (35 E_4^3 +
      37 E_6^2) (3711620489 E_4^9 + 25730000061 E_4^6 E_6^2  \nonumber \\ & \quad +
      25856467947 E_4^3 E_6^4 + 3371520463 E_6^6) +
   108 E_2 E_4 E_6 \Big(5231073695092861 E_4^{15} \nonumber \\
   &\quad +
      93698918450783911 E_4^{12} E_6^2 +
      335105155725269122 E_4^9 E_6^4  \nonumber \\
     &  \quad +   332061543066849710 E_4^6 E_6^6  
    +  87589262461920785 E_4^3 E_6^8 + 4161976813713563 E_6^{10}\Big)\Bigg)  \nonumber
\end{align}
}
\subsection{Modular expressions for $X_{36}$}\label{modX36}
{\small
\begin{align}
F_{H^i}^{(0),inst} &= \sum_{d_1,d_2} P_{ 6(a^1 d_1 + a^2 d_2)-2}^{(0)}(H^i) \Bigg(\frac{q_e^{\frac{1}{2} }}{\eta^{12}}\Bigg)^{a^1 d_1 + a^2 d_2}  \widetilde{Q}_1^{d_1} \widetilde{Q}_2^{d_2}\\
P_{10}(H^1) &= 0\, , \qquad \qquad \qquad \quad P_{10}(H^2) = -3E_4 E_6 \nonumber \, , \\
P_{16}(H^1) & = \frac{3}{2} E_4 + \frac{9}{2} E_4 E_6^2 \,, \quad P_{16}(H^2) = \frac{31}{48}E_4^2 + \frac{113}{48}E_4 E_6^2 \, , \nonumber \\
P_{22}(H^1) & = 0 \, , \qquad \qquad \qquad \quad P_{22}(H^2) = -\frac{17}{32} E_4^2 E_6 - \frac{7}{32} E_4 E_6^3 \, , \nonumber \\
P_{28}(H^1) & = \frac{85}{48}E_4^7 +\frac{3}{8}E_{2} E_4^5 E_6 + \frac{109}{6} E_4^4 E_6^2 + \frac{9}{8} E_2 E_4^2 E_6^3 + \frac{137}{16} E_4 E_6^4 \nonumber\\
P_{28}(H^2) & = \frac{1}{192} E_4 \Big(587 E_4^6 + 103 E_2 E_4^4 E_6 + 5907 E_4^3 E_6^2 +
   329 E_2 E_4 E_6^3 + 2866 E_6^4\Big) \nonumber \\
P_{34}(H^1) & = -\frac{1}{9216}E_4 \Big(48359 E_4^6 E_6 + 161426 E_4^3 E_6^3 + 39047 E_6^5 +
    24 E_2 E_4 \Big(E_4^3 + 3 E_6^2\Big) \Big(31 E_4^3 + 113 E_6^2\Big)\Big) \nonumber \\
 P_{34}(H^2) & = -\frac{1}{110592}
 E_4 \Bigg(208991 E_4^6 E_6 + 755906 E_4^3 E_6^3 + 196319 E_6^5 +
    4 E_2 E_4 \Big(31 E_4^3 + 113 E_6^2\Big)^2\Bigg)   \nonumber \\
F_{H^\circ_i}^{(0),inst} &= \sum_{d_1,d_2} P_{ 6(a^1 d_1 + a^2 d_2)}^{(0)}(H^\circ_i) \Bigg(\frac{q_e^{\frac{1}{2} }}{\eta^{12}}\Bigg)^{a^1 d_1 + a^2 d_2}  \widetilde{Q}_1^{d_1} \widetilde{Q}_2^{d_2} \\
P_{0}(H^\circ_1) & = -24 (b_1 + 1) \sum_{d=1}^{\infty} \frac{\sigma_3(d)}{d^2} q_e^d\, , \quad P_{0} (H^\circ_2) =  -24 (b_2 + 1) \sum_{d=1}^{\infty} \frac{\sigma_3(d)}{d^2} q_e^d\, , \nonumber \\
P_{12}(H^\circ_1)& = \frac{9}{4} E_4^3 + \frac{9}{4} E_6^2\,,  \qquad \quad   P_{12}(H^\circ_2) = \frac{3}{2} E_4^3 + \frac{1}{2} E_2 E_4 E_6 + E_6^2 \, ,\nonumber \\
P_{18}(H^\circ_1) & =  \frac{1}{288}\Big(-31 E_2 E_4^4 - 926 E_4^3 E_6 - 113 E_2 E_4 E_6^2 -
   226 E_6^3\Big) \, ,\nonumber \\
 P_{18}(H^\circ_2) & = \frac{1}{16} \Bigg(-137 E_4^3 E_6 - 47 E_6^3 - 2 E_2 \Big(E_4^4 + 3 E_4 E_6^2\Big)\Bigg) \, ,\nonumber \\ 
P_{24}(H^\circ_1) & = \frac{1}{128} \Bigg(33 E_4^6 - 24 E_2 E_4^4 E_6 +
   2 E_4^2 \Big(-2 E_2^2 + 71 E_4\Big) E_6^2 - 16 E_2 E_4 E_6^3 + 13 E_6^4\Bigg)\, , \nonumber \\
P_{24}(H^\circ_2) & = \frac{1}{384} \Big(51 E_4^6 + 17 E_2 E_4^4 E_6 + 199 E_4^3 E_6^2 +
   7 E_2 E_4 E_6^3 + 14 E_6^4\Big)  \, , \nonumber \\
P_{30}(H^\circ_1) & = \frac{1}{2304}\Bigg(-648 E_2 E_4^7 - E_4^5 \Big(31 E_2^2 + 52747 E_4\Big) E_6 -
 4444 E_2 E_4^4 E_6^2 \nonumber \\
 & \quad - E_4^2 \Big(113 E_2^2 + 105455 E_4\Big) E_6^3 -
 2396 E_2 E_4 E_6^4 - 10422 E_6^5\Bigg) \, ,   \nonumber \\
 P_{30}(H^\circ_2) & = \frac{1}{1152}\Bigg(-386 E_2 E_4^7 - E_4^5 \Big(72 E_2^2 + 32849 E_4\Big) E_6 -
 5002 E_2 E_4^4 E_6^2 \nonumber \\
 & \quad - 24 E_4^2 \Big(9 E_2^2 + 2659 E_4\Big) E_6^3 -
 2100 E_2 E_4 E_6^4 - 6151 E_6^5\Bigg) \, , \nonumber 
 \end{align}
 \begin{align}
 P_{36}(H^\circ_1) & = \frac{1}{1327104}\Bigg(
628895 E_4^9 + 438639 E_2 E_4^7 E_6 + 9743040 E_4^6 E_6^2 +
 1649058 E_2 E_4^4 E_6^3 \nonumber \\
& \quad + 3 E_4^2 \Big(25538 E_2^2 + 3005857 E_4\Big) E_6^4 +
 400623 E_2 E_4 E_6^5 + 392638 E_6^6\Bigg) \, , \nonumber \\
 P_{36}(H^\circ_2) & = \frac{1}{221184}\Big(
333303 E_4^9 + 54875 E_2 E_4^7 E_6 + 5411350 E_4^6 E_6^2 +
 220490 E_2 E_4^4 E_6^3 \nonumber \\
 & \quad + 5526895 E_4^3 E_6^4 + 70235 E_2 E_4 E_6^5 +
 326788 E_6^6\Big) \, . \nonumber 
 \end{align}
 \begin{align}
 F^{(1),inst} &= \sum_{d_1,d_2} P_{ 6(a^1 d_1 + a^2 d_2)}^{(1)} \Bigg(\frac{q_e^{\frac{1}{2} }}{\eta^{12}}\Bigg)^{a^1 d_1 + a^2 d_2}  \widetilde{Q}_1^{d_1} \widetilde{Q}_2^{d_2} \\
 P_{0}^{(1)}& = -2\Big( \frac{\chi}{24} - h_{11}\Big)\sum_{d=1}^{\infty} \frac{\sigma_1(d)}{d}q_e^d \, ,\nonumber \\
P_{12}^{(1)} & = \frac{3}{4} \Big(-6 E_4^3 + 10 E_2 E_4 E_6 - 5 E_6^2\Big) \, , \nonumber \\
P_{18}^{(1)} & = \frac{1}{576} \Big(-2581 E_2 E_4^4 + 9250 E_4^3 E_6 - 8363 E_2 E_4 E_6^2 +
   2990 E_6^3\Big) \, , \nonumber \\
P_{24}^{(1)} & = \frac{1}{6144}\Big(-2565 E_4^6 + 8160 E_2 E_4^4 E_6 - 10454 E_4^3 E_6^2 +
 3360 E_2 E_4 E_6^3 - 805 E_6^4\Big) \nonumber \\ 
 P_{30}^{(1)} & =  \frac{1}{2304}\Bigg(-24891 E_2 E_4^7 + E_4^5 \Big(-4813 E_2^2 + 151361 E_4\Big) E_6 -
 250867 E_2 E_4^4 E_6^2  \nonumber \\
 & \quad + E_4^2 \Big(-15059 E_2^2 + 297124 E_4\Big) E_6^3 -
 121250 E_2 E_4 E_6^4 + 28875 E_6^5\Bigg) \, , \nonumber \\
 P_{36}^{(1)}& =\frac{1}{7962624}\Bigg(-22238425 E_4^9 - 356475921 E_4^6 E_6^2 -
 357370707 E_4^3 E_6^4 - 20115395 E_6^6 + \nonumber \\
& \quad  108 E_2^2 E_4^2 \Big(31 E_4^3 + 113 E_6^2\Big)\Big(577 E_4^3 + 1871 E_6^2\Big) +
 36 E_2\Big(3099607 E_4^7 E_6  \nonumber \\
 & \quad + 10537042 E_4^4 E_6^3 +
    2578903 E_4 E_6^5\Big)\Bigg) \, . \nonumber
\end{align}
}
\label{GeometricInvariants}   

%%xxxxxxxxxxxxxxxxxxxxxxxxxxxxxxxxxxxxxxxxxxxxxxxxxxxxxxxxxxxxxxxxxxxxxxxxxxxxxxxxxxxxxxxxxxx 
\newpage
\label{app:continuation}
\subsection{Analytic continuation data for $X_{24}(1,1,1,1,8,12)$}
We provide the numerical and - as far as we know them - analytic expressions for the continuation matrices $T_c,T_c',T_o,T_o'$ in a Mathematica worksheet on the
webpage \cite{URLsupp}.
Due to their special importance we reproduce here the intersection matrix at $c_1=c_2=0$ as well as
the entries of the continuation matrix to the point $z_e=c_1=0$:
%{\bf Albrecht.  Selected properties of this example can appear in the appendix} 
{\tiny
	\begin{align*}\begin{split}&(T_cT_{c'})^T\eta^{-1}T_cT_{c'}=\kappa\\\cdot&\left(
			\begin{array}{cccccccc}
 0 & 0 & 0 & \frac{2}{3} & 0 & 0 & 0 & 0 \\
 0 & 0 & -6 & \frac{204448}{135} & 0 & 0 & -\frac{60466176}{5} & 0 \\
 0 & -6 & \frac{33392}{15} & -\frac{3091952128}{6075} & 0 & \frac{2519424}{5} & \frac{24428335104}{25} & 0 \\
 \frac{2}{3} & \frac{204448}{135} & -\frac{3091952128}{6075} & \frac{283662214756352}{2460375} & -\frac{46656}{5} & -\frac{3083774976}{25} & -\frac{15768933728256}{125} & 0 \\
 0 & 0 & 0 & -\frac{46656}{5} & 0 & 0 & -\frac{2176782336}{5} & 0 \\
 0 & 0 & \frac{2519424}{5} & -\frac{3083774976}{25} & 0 & \frac{3265173504}{5} & -\frac{12224809598976}{25} & 0 \\
 0 & -\frac{60466176}{5} & \frac{24428335104}{25} & -\frac{15768933728256}{125} & -\frac{2176782336}{5} & -\frac{12224809598976}{25} & \frac{13420960199737344}{125} & 0 \\
 0 & 0 & 0 & 0 & 0 & 0 & 0 & \frac{32}{3} \\
\end{array}
\right)\,,
\end{split}
\end{align*}
}
where
\begin{align}
\kappa=\frac{1}{1572864\pi^4}\,.
\end{align}
{\tiny
\begin{align*}
\begin{split}
f_{1,1}=&\frac{-2916 \pi ^{12} r_4^4+10260 \pi ^6 r_4^2-27 \pi ^4 r_5 r_4+144 r_1^2-32 r_2^2-7129}{1152}\\
f_{2,1}=&\frac13 (6r_1+4ir_2+3)\\
f_{3,1}=&\frac{1}{384} \left(216 \pi ^6 r_4^2+48 r_1+32 i r_2+3 i \pi  r_5-404\right)\\
f_{7,1}=&-\frac{1}{12}i(2r_2-9\pi^3r_4)\\
f_{5,1}=&\frac{1}{48} \left(-54 \pi ^6 r_4^2-36 i \pi ^3 r_4-12 r_1+8 i r_2+101\right)\\
f_{1,2}=&\frac{1}{24} \left(3 r_1 r_3-\sqrt{2} r_2 r_3+384 \pi ^2 r_4-24 r_5\right)\\
f_{3,2}=&\frac{-2304 i \pi ^6 r_4^2+768 \pi ^3 r_4+i \sqrt{2} \pi  r_3+\pi  r_3+3968 i}{16 \pi }\\
f_{5,2}=&-\frac{-i \sqrt{2} \pi  r_3+\pi  r_3+768 \pi ^3 r_4+256 i}{8 \pi }\\
f_{1,3}=&-\left(116640 \pi ^{16} r_3 r_4^4-410400 \pi ^{10} r_3 r_4^2-207360 \pi ^8 r_3 r_4^2-71424 \pi ^6 r_3 r_4+276480 \pi ^4 r_3 r_4+1080 \pi ^8 r_3 r_5 r_4-45 \pi ^4 r_1 r_3^2\right.\\
&\left.+15 \sqrt{2} \pi ^4 r_2 r_3^2+11796480 r_1+3932160 \sqrt{2} r_2-5760 \pi ^4 r_1^2 r_3+1280 \pi ^4 r_2^2 r_3\right.\\
&\left.+285160 \pi ^4 r_3+349440 \pi ^2 r_3+4464 \pi ^4 r_3 r_5\right)/(768 \pi ^4 r_3)\\
f_{2,3}=&\frac{5 \left(3 i \sqrt{2} \pi ^4 r_3^2+3 \pi ^4 r_3^2+768 \pi ^4 r_1 r_3+512 i \pi ^4 r_2 r_3+384 \pi ^4 r_3+786432 i \sqrt{2}-786432\right)}{32 \pi ^4 r_3}\\
f_{3,3}=&\left(15 i \sqrt{2} \pi ^4 r_3^2+15 \pi ^4 r_3^2+17280 \pi ^{10} r_4^2 r_3-428544 i \pi ^9 r_4^2 r_3+3840 \pi ^4 r_1 r_3+2560 i \pi ^4 r_2 r_3+142848 \pi ^6 r_4 r_3-92160 i \pi ^5 r_4 r_3\right.\\
&\left.+240 i \pi ^5 r_5 r_3-32320 \pi ^4 r_3+738048 i \pi ^3 r_3+15360 \pi ^2 r_3+61440 i \pi  r_3+3932160 i \sqrt{2}-3932160\right)/(512 \pi ^4 r_3)\\
f_{4,3}=&\frac{15 \left(3 i \sqrt{2} \pi ^4 r_3^2+2 \pi ^4 r_3^2+512 \pi ^4 r_1 r_3+512 i \pi ^4 r_2 r_3+512 \pi ^4 r_3+786432 i \sqrt{2}-524288\right)}{128 \pi ^4 r_3}\\
f_{5,3}=&-\left(-15 i \sqrt{2} \pi ^4 r_3^2+15 \pi ^4 r_3^2+17280 \pi ^{10} r_4^2 r_3+3840 \pi ^4 r_1 r_3-2560 i \pi ^4 r_2 r_3+11520 i \pi ^7 r_4 r_3+142848 \pi ^6 r_4 r_3-32320 \pi ^4 r_3\right.\\
&\left.+47616 i \pi ^3 r_3+15360 \pi ^2 r_3-3932160 i \sqrt{2}-3932160\right)/(256 \pi ^4 r_3)\\
f_{6,3}=&\frac{5 i \left(3 \sqrt{2} \pi ^4 r_3^2+512 \pi ^4 r_2 r_3+786432 \sqrt{2}\right)}{64 \pi ^4 r_3}\\
f_{7,3}=&\frac{i \left(-15 \sqrt{2} \pi ^4 r_3^2-2560 \pi ^4 r_2 r_3+11520 \pi ^7 r_4 r_3+47616 \pi ^3 r_3-3932160 \sqrt{2}\right)}{256 \pi ^4 r_3}\\
f_{1,4}=&\frac{1}{64}(16\pi^2r_4-r_5)\\
f_{3,4}=&\frac{-18i\pi^6r_4^2+6\pi^3r_4+32i}{8\pi}\\
r_1=&0.0333238838392332919265429398082\\r_2=&-1.\
29219644630091977480074761037\\r_3=&74.0860643209298158454123721134\\\
r_4=&-0.00948778220735050311547607017424\\r_5=&122.\
032462442689559692241449686
\end{split}
\end{align*}
}
\addcontentsline{toc}{section}{References}
\bibliographystyle{utphys}
\bibliography{act}
\end{document}